\documentclass[10pt]{article}

\input{header}

\title{Graph fission and cross-validation}
\author{
James Leiner\textsuperscript{1} , Aaditya Ramdas\textsuperscript{1,2}
}
\date{}

\begin{document}
\maketitle

\begin{center}
\begin{tabular}{l}
\textsuperscript{1}Department of Statistics and Data Science, Carnegie Mellon University \\
\textsuperscript{2}Machine Learning Department, Carnegie Mellon University \\

\end{tabular}

\texttt{\{jleiner,aramdas\}@stat.cmu.edu}\\
\end{center}

\begin{center}
\today
\end{center}

\begin{abstract}
We introduce a technique called graph fission which takes in a graph which potentially contains only one observation per node (whose distribution lies in a known class) and produces two (or more) independent graphs with the same node/edge set in a way that splits the original graph's information amongst them in any desired proportion. Our proposal builds on data fission/thinning, a method that uses external randomization to create independent copies of an unstructured dataset. 
We extend this idea to the graph setting where there may be  latent structure between observations. We demonstrate the utility of this framework via two applications: inference after structural trend estimation on graphs and a model selection procedure we term ``graph cross-validation''.
\end{abstract}

\section{Introduction}
Sample splitting, where an analyst divides a portion of data to train a model and the remaining portion of data to validate it is a ubiquitously used tool by statisticians. Unfortunately, this approach is typically appropriate only in settings with repeated $i.i.d.$ observations, with few exceptions. In cases where the available data is dependent or not identically distributed, for instance time series or fixed-design regression, sample splitting is generally not practical or easily interpretable. 
Nonetheless, sample splitting strategies are often still used due to a lack of alternatives. This often leads to an analysis pipeline that is heuristic and does not come with theoretical guarantees. 

Building on an idea for post-selection inference in linear models~\citep{tian2018selective,rasines2021splitting}, recent work by \citet{data_fission} proposed a general  solution to this problem through use of external randomization to create synthetic copies of the data that are  $i.i.d.$ by construction and each of which contain a portion of the information contained in the original dataset. Such a construction does not always work: it is only feasible if the distribution of the data lies in certain known classes. While~\citet{data_fission} applied to to varied contexts like estimation after multiple testing and post-selection inference for trend filtering, follow-up work \citep{neufeld_biostats,neufeld2023data,dharamshi2023generalized} has provided improved constructions in several settings and applied it to new settings like latent variable estimation. 

In this paper, we will apply these procedures to the graph setting and demonstrate their efficacy through two practical applications: cross-validation for graph-valued data and inference after structural trend estimation over graphs. Graph-valued data presents unique challenges because in some settings, only a single graph is observed and there is a learning or inference problem on the graph, where one believes that the graph structure captures the structure in either the signal or the noise (i.e.\ either the signal is smooth over the graph, or the graph captures the correlation structure of the noise). 
By extending the aforementioned ``fissioning'' idea to graphs, an analyst can gain access to techniques that are not generally available in settings without repeated $i.i.d.$ observations, including cross-validation to estimate out-of-sample risk and the creation of independent training, validation and test graphs (or selection and inference graphs) so that the analyst may explore different modeling techniques without violating error control on downstream inferential procedures via ``double dipping''.

\paragraph{Contributions.} Our main contributions are:
\begin{itemize}
\item When the distribution of errors is known and falls into the ``convolution-closed'' class \citep{joe_conv}, we demonstrate how data fissioning \citep{data_fission} and thinning \citep{neufeld2023data} methods can enable cross-validation on graphs for hyperparameter tuning. We demonstrate the utility of this procedure on trend estimation problems, but note that it applies generically to procedures that requires hyperparameter tuning over graphs, such as the training of graph neural networks.
\item We extend the results of \citet{data_fission,neufeld2023data} to enable inference for structural trend estimation on graphs in the Gaussian setting, in the presence of increasing dimensions and unknown error variance. Existing results all rely heavily on known specification of errors or a low-dimensional setting that ensures consistent estimators of the error variance are available. We provide a result (\cref{thm:robust-ci}) that enables graph fission in the presence of unknown error variance in the Gaussian case. This allows for the application of these techniques in real-world settings where the variance is not known a priori. 
\end{itemize}

\paragraph{Paper outline.} 
In \cref{sec:methodology}, we review decompositions for fissioning a single dataset into $m$ independent copies and provide examples of how this method can be used in the context of graph-valued data. We also discuss common methods for estimating a structural trend over a graph. In \cref{sec:cross-validation}, we build on these techniques by introducing graph cross-validation, which we then use to tune hyperparameters used in trend estimation. In \cref{sec:inference_after_trend_estimation}, we demonstrate a second application of graph fission: the creation of valid confidence intervals after model selection on a graph. In \cref{sec:real_data_example}, we apply these methods to real data by constructing confidence intervals on taxicab usage in New York City. We conclude in \cref{sec:conclude}.

\section{Methodology} \label{sec:methodology}
Let $\mathcal{G} = (V,E,Y)$ be a graph with a known vertex ($V$) and edge ($E$) set and a set of observations $Y = (y_{1},...,y_{n}) \in \mathbb{R}^{n}$ over the vertices. We use a standard nonparametric regression framework, where 
$$ y_{i} = \mu_{i} + \epsilon_{i},$$
$\mu_{i} = E[y_{i}]$, and $\epsilon_{i}$ is a $0$ mean random variable. Denote $\mu = (\mu_{1},...,\mu_{n})$ and $\epsilon = (\epsilon_{1},...,\epsilon_{n})$. 



\subsection{Decomposition Rules} \label{sec:decomposition}
We aim to create $m$ new synthetic copies of $\mathcal{G}$, which we denote as $\mathcal{G}_{1},...,\mathcal{G}_{m}$ with corresponding observations labeled as $Y^{\mathcal{G}_{1}},...,Y^{\mathcal{G}_{m}}$. We require the synthetic graphs to have the following properties: 
\begin{enumerate}
    \item $\mathcal{G}_{i}$ has the same non-random structure (i.e. $V$ and $E$). Furthermore, $\mathbb{E}[Y^{\mathcal{G}_{j}}] = h_{1}(\mu)$ for all $j \in [m]$ and some known deterministic function $h_{1}$.
    \item Taken together, the individual datasets recover the original data $Y$ in the sense that there exists a known deterministic function $h_{2}$ such that $\mathcal{G} = h_{2}\left(\mathcal{G}_{1},...,\mathcal{G}_{m}\right)$, and you can not recover $\mathcal{G}$ from any strict subset of graphs $\mathcal{G}_{1},...,\mathcal{G}_{m}$
    \item The information contained in $Y$ is divided across $\mathcal{G}_{1},...,\mathcal{G}_{m}$ in any proportion desired.  
\end{enumerate}

\begin{remark} \label{rmk:P2}
At this stage, we also note the existence of an alternative less stringent set of requirements. Following the terminology of \cite{data_fission}, we call this the \textbf{P2} regime. Here, we only require the creation of two synthetic copies of these graphs $\mathcal{G}_{1},\mathcal{G}_{2}$. The properties that they must fulfill are: 
\begin{itemize}
    \item The law of $Y^{\mathcal{G}_{2}} | Y^{\mathcal{G}_{1}}$ is known and tractable. 
    \item There exists a function $h$ such that $\mathcal{G} = h(\mathcal{G}_{1},\mathcal{G}_{2})$. 
\end{itemize}
Although we will not focus on this idea throughout much of the paper, we will revisit it in \cref{sec:inference_after_trend_estimation}, as it is a key ingredient for \cref{thm:robust-ci}. 
\end{remark}

Following the terminology of ``data fission''~\citep{data_fission}, we call the above task ``graph fission''.
Instead of using the preceding paper's techniques, we instead employ the decomposition rules of \cite{neufeld2023data} which provide an algorithm for splitting a random variable into $m$ independent copies when the distribution of that variable is \emph{convolution-closed}.
\begin{definition}[\cite{joe_conv}]
Let $F_{\theta}$ be a distribution indexed by a parameter $\theta$ in parameter space $\Theta$. Drawing $X' \sim F_{\theta_{1}}$ and $X'' \sim F_{\theta_{2}}$ independently, if $X' + X'' \sim F_{\theta_{1} + \theta_{2}}$ whenever $\theta_{1} + \theta_{2} \in \Theta$ then $F_{\theta}$ is \emph{convolution-closed} in the parameter $\theta$.  
\end{definition}
Note that many distributions encountered in data analysis including Gaussian, Poisson, binomial,  and negative binomial are convolution-closed. 

A property of many convolution-closed distributions is that when $X_{1},...,X_{m}$ are each drawn from the same family of distributions, then the joint density of $(X_{1},...,X_{m}) | \sum_{i=1}^{m} X_{i} = x$ is tractable.  Hence, $m$ synthetic samples can then be generated from a single sample $x$ by drawing from this joint distribution, conditioning on their total sum adding up to $x$. This intuition is formalized below.

\begin{figure}
\centering
    \includegraphics[width=0.8\linewidth]{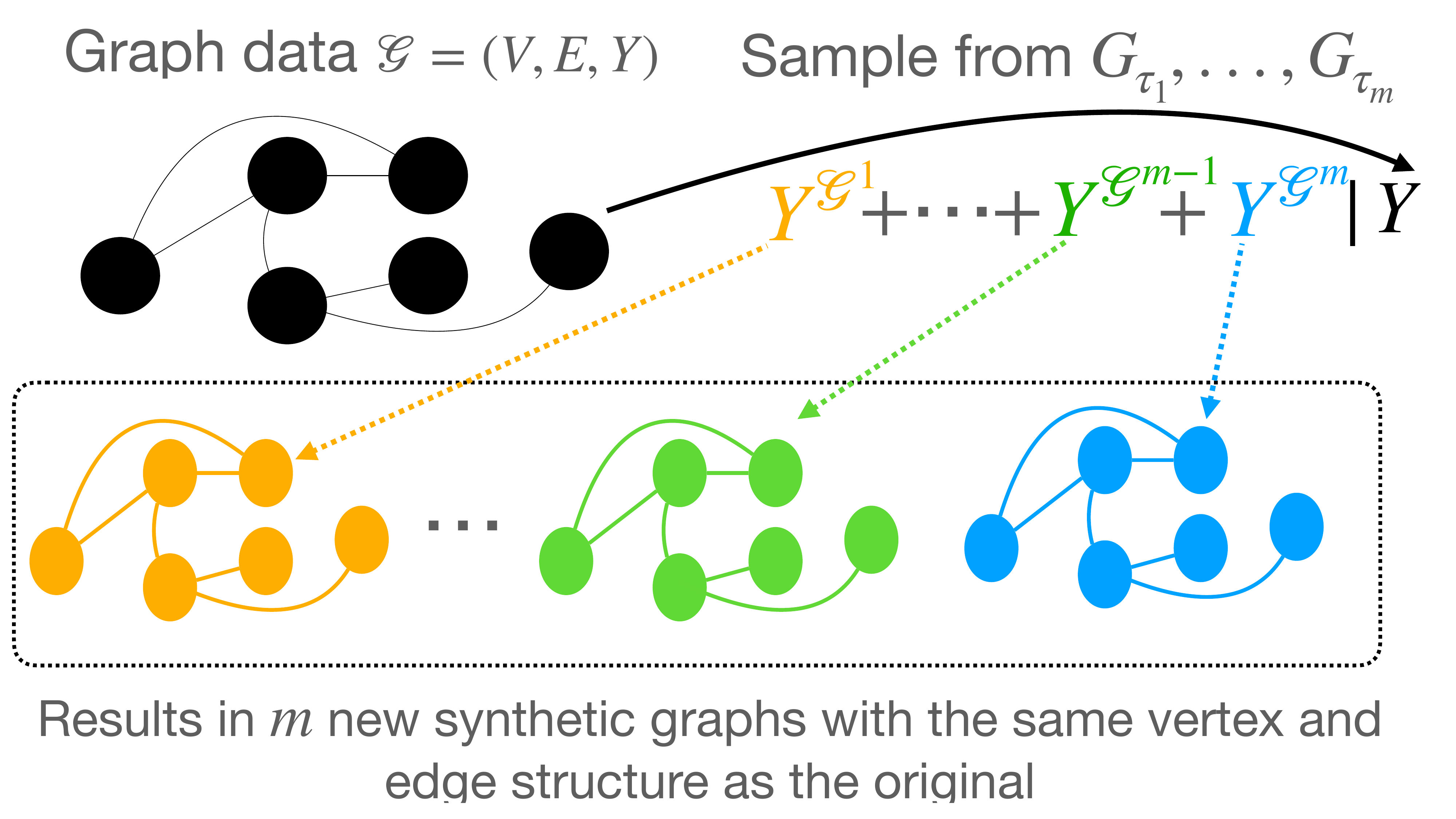}
    \caption{Graphical illustration of \cref{fact:data_thinning}.}

    \label{fig:illustration_thinning}
\end{figure}

\begin{fact}[Theorem 2 of \cite{neufeld2023data}]
Assume $X \sim F_{\theta}$ for a convolution-closed distribution in parameter $\theta \in \Theta$. Choose $\tau_{1},...,\tau_{m}$ such that $\sum_{j=1}^{m} \tau_{j} =1$ and $\tau_{j}\theta \in \Theta$. Let $G_{\theta_{1},...,\theta_{m}}$ be the joint distribution of $(X_{1},...,X_{m}) | \sum_{i=1}^{m} X_{i} = X$ when $X_{i}$ are drawn independently from $F_{\theta_{i}}$. If $X_{1},...,X_{m} \sim G_{\tau_{1}\theta,...,\tau_{m}\theta}$,then the following holds: (i) $X_{i} \sim F_{\tau_i \theta}$;
(ii) $X_{1},...,X_{m}$ are mutually independent; (iii)  $\sum_{i=1}^{m} X_{i} = X$; (iv) whenever $F_{\theta}$ has finite first moment, $\mathbb{E}\left[X_{j} \right] = \theta_{j}\mathbb{E} \left[ X\right]$.
\label{fact:data_thinning}
\end{fact}
\cref{fact:data_thinning} can be applied whenever the joint distribution of $Y$ is convolution-closed (a special case of this being when $y_{i}$ are individually convolution-closed and mutually independent), and the analyst is able to draw from the distribution $G_{\tau_{1},...,\tau_{m}}$. See \cref{fig:illustration_thinning} for an illustration of this procedure. As a first example, we will use \cref{fact:data_thinning} to fission a graph with Gaussian errors.

\begin{example}[Gaussian graph]
Assume $y_{i} \sim N(\mu_{i}, \sigma^{2})$. We draw $y_{i}^{\mathcal{G}_{1}},...,y_{i}^{\mathcal{G}_{m}}$ from the distribution
$$N \left(\begin{bmatrix}
           y_{i} \\
           \vdots \\
           y_{i}
         \end{bmatrix}, \sigma^{2} \begin{bmatrix}
            (m-1) & -1& \hdots & -1 \\
            -1 & (m-1)& \hdots &\vdots \\
            \vdots & & \ddots & \\
            -1 & -1 &\hdots & (m-1)
            \end{bmatrix} \right).$$
Marginally, $y_{i}^{\mathcal{G}_{j}} \sim N(\mu_{i},m\sigma^{2})$, $j \in [m]$, all mutually independent. Note that this procedure splits the information evenly across all graphs, because the Fisher information $\mathcal{I}_{\mathcal{G}^{j}}( \mu_{i}) = \frac{1}{m \sigma^{2}}$ for all $j$. \label{example:gaussians}
\end{example}

\begin{example}[Gaussian graph with correlated errors]
The preceding example can be generalized to the correlated Gaussian case, albeit with a more complicated decomposition strategy. Assume $Y \sim N(\mu, \Sigma)$, where $\Sigma$ is any (known) covariance matrix. Then draw 
$Y^{\mathcal{G}_{1}} \mid Y  \sim N(Y,  (m-1) \Sigma  )$ and $Y^{\mathcal{G}_{-1}} :=  \frac{m}{m-1}Y - \frac{1}{m-1}Y^{(1)}$. Marginally, $Y^{\mathcal{G}_{1}}\sim N(\mu, m\Sigma)$ and $Y^{\mathcal{G}_{-11}} \sim N(\mu, \frac{m}{m-1} \Sigma)$. Continuing to draw $Y^{\mathcal{G}_{j+1}} \mid Y^{\mathcal{G}_{-j}}  \sim N( Y^{\mathcal{G}_{-j}}, (m-j-1)\frac{m}{m-j} \Sigma)$ with  $Y^{\mathcal{G}_{-j-1}}  := \frac{m}{m-j} Y^{\mathcal{G}_{-j}}  - \frac{1}{m-j}Y^{\mathcal{G}_{j+1}}$ and proceeding $m$ times will result in $ Y^{\mathcal{G}_{j}} \stackrel{i.i.d.}{\sim}  N(\mu, m \Sigma)$.
\end{example}

One drawback is that the covariance matrix must be known prior to fissioning. We defer discussion of the case of unknown $\sigma$ to \cref{sec:inference_after_trend_estimation}. 
Not all distributions will have this issue---eg: Poisson errors can be fissioned without needing to estimate an unknown parameter. 


\begin{example}[Poisson graph]Assume $y_{i} \sim \text{Pois}(\mu_{i})$. We can draw a new vector $y_{i}^{\mathcal{G}_{1}},...,y_{i}^{\mathcal{G}_{m}}$ as $\text{Multinomial}\left(y_{i},\left(\frac{1}{m},...,\frac{1}{m}\right)\right)$. Then each component $y_{i}^{\mathcal{G}_{j}} \sim \text{Pois}\left(\frac{\mu_{i}}{m} \right)$. We again note that this divides the Fisher information evenly across each graph. 
\end{example}

Note that in the preceding examples, we index over both the nodes ($i\in [n]$) and synthetic samples at each node ($j \in [m]$). For clarity, we keep this notation consistent throughout the paper, with $j$ always indexing over synthetic samples and $i$ indexing over nodes. 

Both of these examples, as well as several more, can be found in \cite{data_fission,neufeld2023data}, but the latter paper has a more unified treatment.

\subsection{Structural Trend Estimation on Graphs} \label{sec:trend_estimation}
As a unifying example to work with across the paper, we consider estimating a structural trend over $\mu$ by solving an optimization problem of the form,
\begin{equation}
    \hat{\beta} := \argmin_{\beta \in \mathbb{R}^{n}} \underbrace{\ell(Y,\beta)}_{\text{Loss}} + \underbrace{D(\beta)}_{\text{Penalty}}.
    \label{eqn:opt}
\end{equation}
The loss can be any convex function, but we will focus on square loss $\ell(Y,\beta) := \frac{1}{n} \norm{Y - \beta}_{2}^{2}$ for continuous-valued graph data and Poisson loss $\ell(Y,\beta) := \frac{1}{n}\sum_{i=1}^{n} \left( -y_{i} \beta_{i} + \exp(\beta_i) \right) $ for count data. 

In order to estimate a smooth trend that aligns with the graph structure given by the vertex and edge set, most approaches will add a regularization term that encourages smoothing over adjacent nodes \citep{kondor_lafferty,pmlr-v31-sharpnack13a,graph_tf}. Graph trend filtering \citep{graph_tf} accomplishes this by introducing a graph difference operator defined as $\Delta^{(1)} \in \{-1,0,1 \}^{n \times p}$, which contain a single row for each of the $p$ edges in the graph. The row corresponding to a particular edge $e_{v} = (i,j)$ is  
$$ \Delta_{v}^{(1)} =(0, \ldots \underset{\substack{\uparrow \\ i}}{-1}, \ldots  \underset{\substack{\uparrow \\ j}}{1}, \ldots 0),$$
where the position of $1$ and $-1$ is arbitrary. Setting the penalty $D(\beta) := \lambda \norm{\Delta^{(1)} \beta}_{1}$ penalizes any differences between values of $\hat{\beta}$ at adjacent nodes, leading to a piecewise constant solution across connected components. Recursively applying this operator yields,
$$\Delta^{(k+1)}= \begin{cases}\left(\Delta^{(1)}\right)^{\top} \Delta^{(k)}=L^{\frac{k+1}{2}} & \text { for odd } k \\ \Delta^{(1)} \Delta^{(k)}=\Delta^{(1)} L^{\frac{k}{2}} & \text { for even } k\end{cases},$$
where $L$ denotes the graph Laplacian. The corresponding penalty term $\norm{\Delta^{(k+1)} \beta}_{1}$ penalizes higher order graphs differences in an analogous fashion. For instance, $k=1$ enforces a piecewise linear structure across connected components, $k=2$ enforces a piecewise quadratic structure, and so on. 

\begin{figure*}
\includegraphics[width=0.19\textwidth]{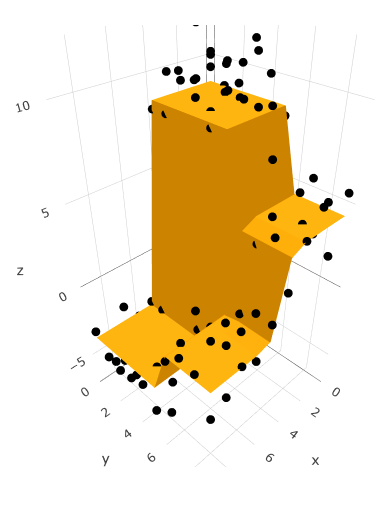}
\includegraphics[width=0.19\textwidth]{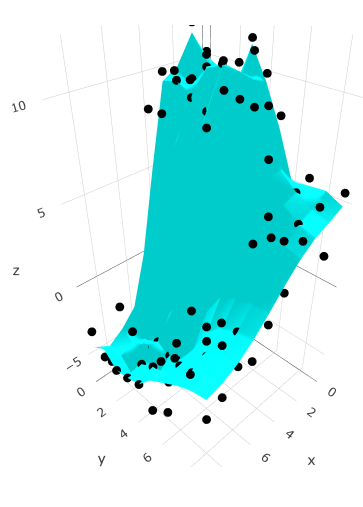}
\includegraphics[width=0.19\textwidth]{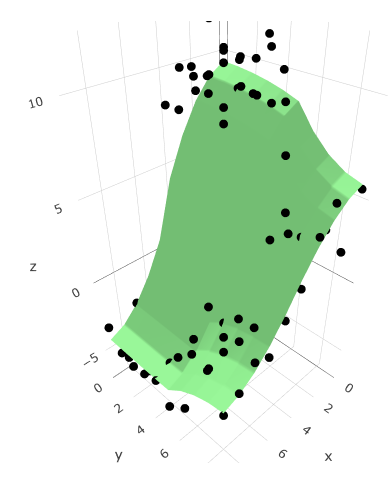}
\includegraphics[width=0.19\textwidth]{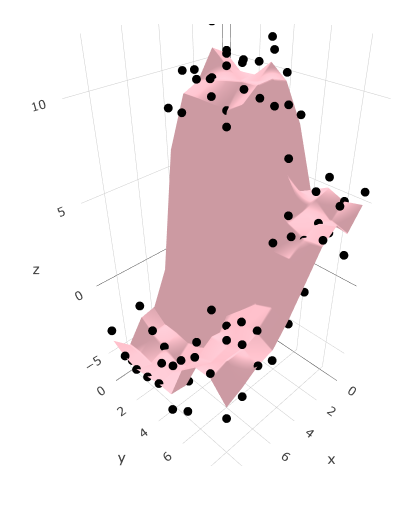}
\includegraphics[width=0.19\textwidth]{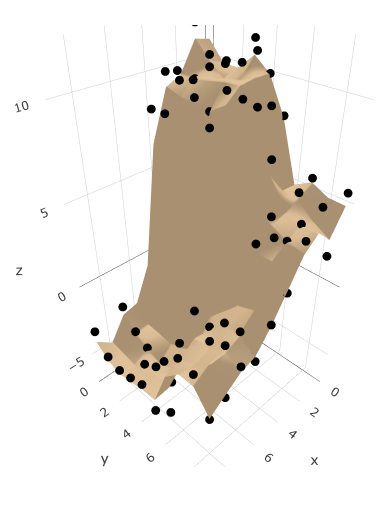}
\caption{Example of synthetic data points and corresponding structural trend solution $\hat{\beta}$ when fit using square loss and a variety of penalties. From left to right: piecewise constant $\left( \norm{\Delta^{(1)}\beta}_{1}\right)$, linear $\left( \norm{\Delta^{(2)}\beta}_{1}\right)$, quadratic $\left( \norm{\Delta^{(3)}\beta}_{1}\right)$, ridge $\left( \norm{\Delta^{(1)}\beta}_{2}^{2}\right)$, and elastic net$\left( \alpha \norm{\Delta^{(1)}\beta}_{1} + (1- \alpha) \norm{\Delta^{(1)}\beta}_{2}^{2}\right)$.} 
\label{fig:example_penalties}
\end{figure*}

Other ways of constructing the penalty term include using an $L_{2}$ penalty term (i.e. $D(\beta) :=\lambda \norm{\Delta^{(k+1)} \beta}_{2}^{2}$)  instead of $L_{1}$, which is equivalent to graph Laplacian smoothing \citep{laplacian_smoothing}, and  combining the two to create an elastic net penalty term by setting $D(\beta) := \lambda_{1} \norm{\Delta^{(k+1)} \beta}_{1} + \lambda_{2}\norm{\Delta^{(k+1)} \beta}_{2}^{2}$. 

In the case where $L_{2}$ regularization is used alongside square loss, the solution can be computed in closed form as $\hat{\beta} = \left(I + n\lambda\left(\Delta^{(k+1)}\right)^{T} \Delta^{(k+1)} \right)^{-1}Y $. When $L_{1}$ regularization is used alongside square loss, \cite{dual_path_algorithms} present efficient algorithms to compute $\hat{\beta}$ along any solution path of $\lambda$ which are currently implemented in the R package \textsc{genlasso}. 

The solutions corresponding to other convex loss functions can be computed through standard optimization techniques such as gradient descent, though designing algorithms that can compute these solution paths most efficiently is an open area of investigation. 

As an illustration, we generate data on a $10 \times 10$ grid. Nodes at Manhattan distance 1 share an edge, resulting in a graph with $100$ nodes and $180$ edges. We then estimate the structural trend using square loss and a variety of different penalties in \cref{fig:example_penalties}. In the case of $L_{1}$ penalties, the fitted solutions become piecewise polynomials. For the $L_{2}$ and elastic net penalties, fewer components are chosen to be exactly $0$ so the structural trend tends to pick up more local variation.

\section{Graph Cross-Validation}
\label{sec:cross-validation}
Across all methods for estimating the structural trends over graphs, the training of hyperparameters is a consistent challenge. When $i.i.d.$ data is observed, for instance in random-design linear regression, the most common method for tuning $\lambda$ is via cross-validation which does not have a direct analog in the graph setting where data is not identically distributed. Some existing methods for performing cross-validation on structured data \citep{structured_cv,cv_hmm} can be applied to the graph setting, but they rely on an assumption that the structural trend does not vary substantially within the held-out set which may not be reasonable for datasets with substantial variation across nodes. We will not need any such assumption, as discussed below.

Another alternative to cross-validation 
is to choose $\lambda$ is to minimize Stein's Unbiased Risk Estimate (SURE) \citep{10.1214/aos/1176345632}, but this approach comes with its own set of drawbacks. SURE is an unbiased estimate of the risk only in the case of Gaussian data (though data fission was recently used to extend SURE to the Poisson case~\citep{oliveira2022poisson}). Moreover, it requires knowledge of the effective number of degrees of freedom of the estimator which may not be readily available in all instances. Lastly, although SURE and cross-validation are asymptotically equivalent when the data is Gaussian, cross-validation often has superior finite sample performance.

Motivated by these concerns, we apply the methodology of \cref{sec:decomposition} to construct an analog of cross-validation in the graph setting. 

\begin{assumption}
Assume that the distribution of $Y$ is convolution-closed and follows distribution $F_{\theta}$.
\label{assumption:conv_close}
\end{assumption}
Under \cref{assumption:conv_close}, we are able to generate $m$ new independent graphs $\mathcal{G}_{1},...,\mathcal{G}_{m}$ such that 
$\mathbb{E}\left( Y^{\mathcal{G}_{j}} \right) = \frac{1}{m} \mu$ and $Y^{\mathcal{G}_{j}} \sim F_{\frac{\theta}{m}}$ for all $j \in [m]$ using \cref{fact:data_thinning}. To perform cross-validation, we average $m-1$ of these graphs together and leave the remaining held out graph  for testing. Denote $Y^{\mathcal{G}_{-j}} := \sum_{j \ne i} Y^{\mathcal{G}_{j}}$. By construction,
$$ Y^{\mathcal{G}_{-j}} \sim F_{\theta \frac{m-1}{m}} \text{  and } Y^{\mathcal{G}_{j}} \sim F_{\theta \frac{1}{m}}.$$ 


We can therefore use $\mathcal{G}_{-j}$ to estimate $\theta$ and evaluate this estimate using the held out graph $\mathcal{G}_{j}$ to get an unbiased estimate of the risk of the procedure. Repeating this process $m$ times for each $j \in [m]$ mimics the process of $m$-fold cross-validation. 
\begin{remark}
If $F$ is a location-scale distribution, we can rescale $Y^{\mathcal{G}_{-j}}$ and $Y^{\mathcal{G}_{j}}$ so their distributions are functions of the same parameter (e.g, \cref{example:gaussians}). 
Otherwise, the analyst will need to be careful to scale $\hat{\theta}$ appropriately when out-of-the-box estimation procedures are used so evaluation on the test set is comparable with the training data. 
\end{remark}

\paragraph{Gaussian Data with Unknown Variance.} If $y_{i} \sim N(\mu_{i},\sigma^{2})$ and $\sigma^{2}$ is known, then \cref{fact:data_thinning} can be applied directly. When $\sigma^{2}$ is unknown, it will need to be estimated. Estimating the error variance in high dimensional regression problems is difficult. Most estimators are only provably consistent under assumptions that are not verifiable in practice \citep{fan_errors} and rely on techniques like cross-validation \citep{hd_var_estimates} which we do not have access to. Nonetheless, a heuristic that works empirically is to estimate
\begin{equation} \label{eqn:variance_estimate}
\hat{\sigma}^{2} := \frac{1}{n - \text{df}(\hat{\beta}_{\lambda})} \norm{Y - \hat{\beta}_{\lambda}}_{2}^{2},
\end{equation}
where $\hat{\beta}_{\lambda}$ is fit using a fixed $\lambda$ lasso penalty. Since $\lambda$ is typically chosen either through cross-validation or by minimizing SURE (which pre-supposes knowledge of $\sigma^{2}$), we unfortunately have to use a pre-determined version of $\lambda$ to make this selection. We choose $\lambda = \sqrt{\frac{\log{n}}{n}}$ since many results require that $\lambda$ grow at this rate to ensure consistency \citep{bien_errorvariance}.

\paragraph{Simulation.} We apply this method to the problem of estimating an optimal $\lambda$ for structural smoothing problems as described in \cref{sec:methodology}. We again use a node set aligned in a grid such that each vertex $v \in [1,10] \times [1,10]$ where nodes at Manhattan distance 1 are connected. A ground truth trend $\mu$ is constructed by randomly choosing a percentage of nodes to be active nodes that permit a structural change relative to adjacent nodes. We then generate piecewise polynomials over the connected components of the inactive set of nodes and draw observations as $y_{i} \sim N(\mu_{i}, 1)$. 

The structural trend $\hat{\beta}$ is estimated using square loss and a penalty term 
as described in \cref{sec:trend_estimation}. When using graph fission to select $\lambda$, we consider a rule which picks the $\lambda$ that minimizes the average test error (over each of the $m$ folds) as well as the so called ``one-standard error'' rule which picks the largest value of $\lambda$ (i.e. the most parsimonious model) which falls within one standard deviation of the minimum error.

As a point of comparison, we compare graph cross-validation with ``ordinary'' cross-validation in the form of structured cross-validation proposals \citep{structured_cv,cv_hmm}, adjusted to the graph setting. These amount to selecting a subset of nodes $\mathcal{I} \subseteq V$ to hold out for evaluation. We then estimate a structural trend $\hat{\beta}_{-\mathcal{I}}$ by ignoring the data in $\mathcal{I}$, and define the trend $\hat{\beta}_{\mathcal{I}}$ in the holdout set to be the average of adjacent nodes in the training set for each point. This amounts to assuming that all the nodes in the held-out set are inactive so the structural trend can be interpolated from the training set. This approximation makes the most sense when there is minimal structural change within $\mathcal{I}$, which corresponds to a smooth trend over the graph with few breakpoints.

\begin{figure*}  
    \centering
\includegraphics[width=0.48\textwidth]{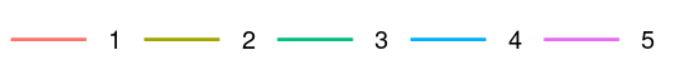}
\includegraphics[width=0.28\textwidth]{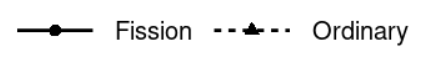} \\
    \begin{subfigure}[t]{0.48\linewidth}
         \centering
         \includegraphics[width=0.49\textwidth]{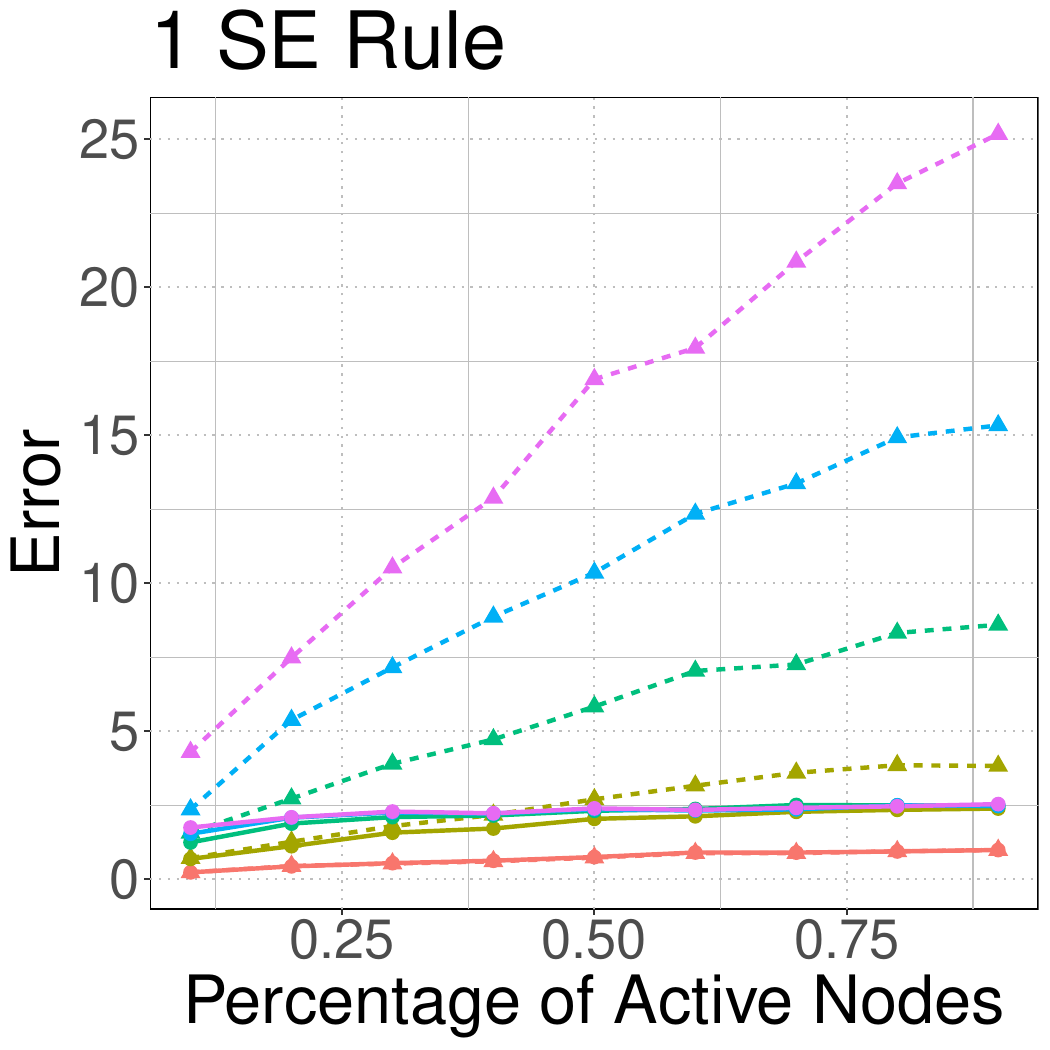}
        \includegraphics[width=0.49\textwidth]{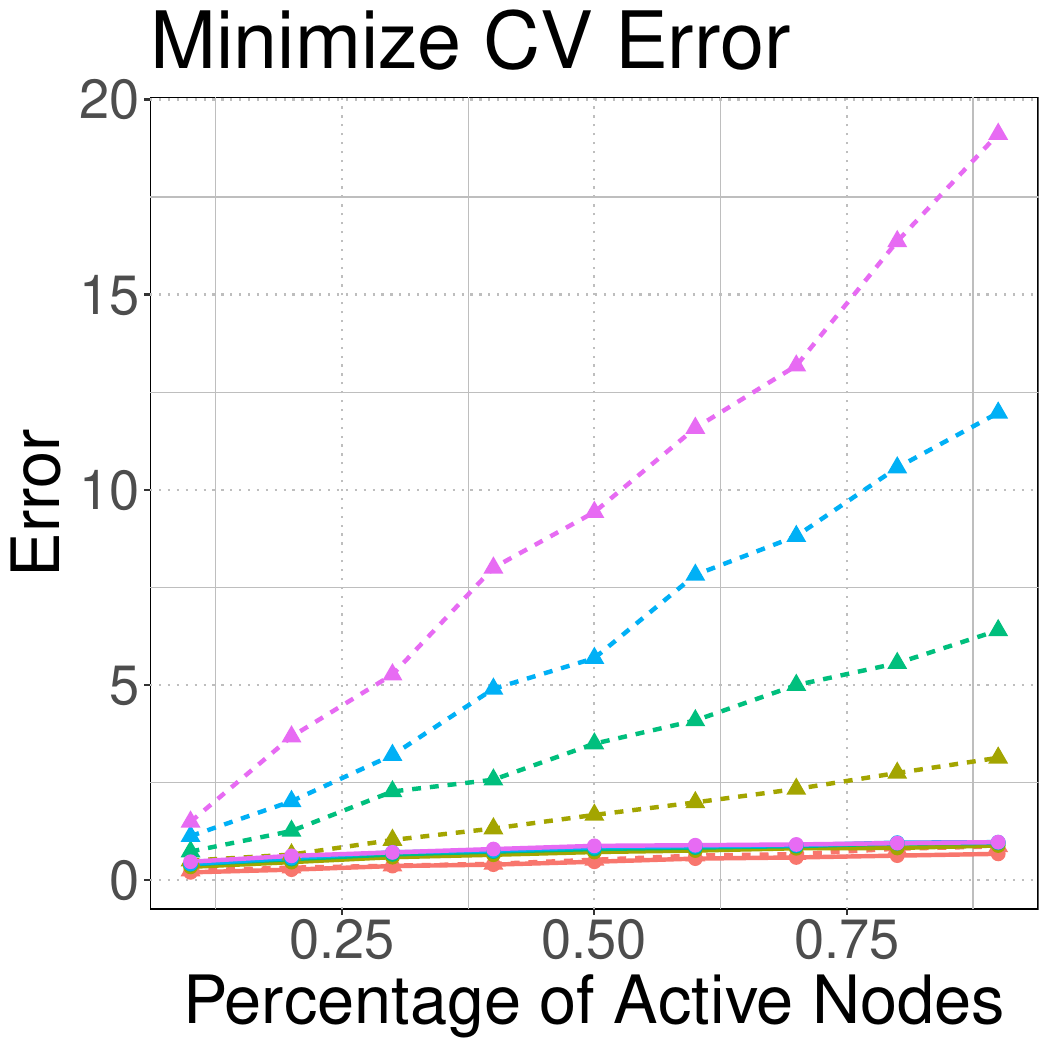}
        \caption{Piecewise constant trend filtering ($k=0$)}
     \end{subfigure}
    \begin{subfigure}[t]{0.48\linewidth}
         \centering
         \includegraphics[width=0.49\textwidth]{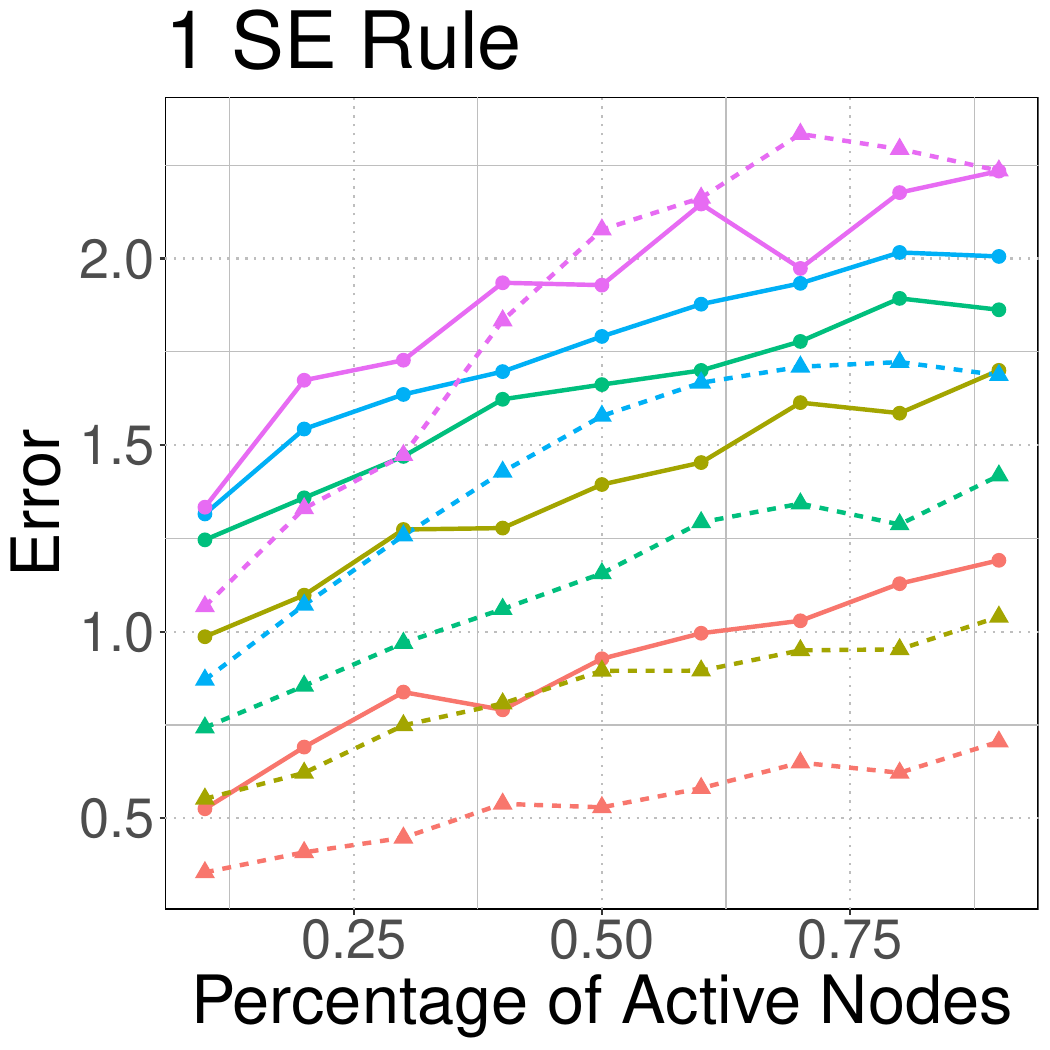}
        \includegraphics[width=0.49\textwidth]{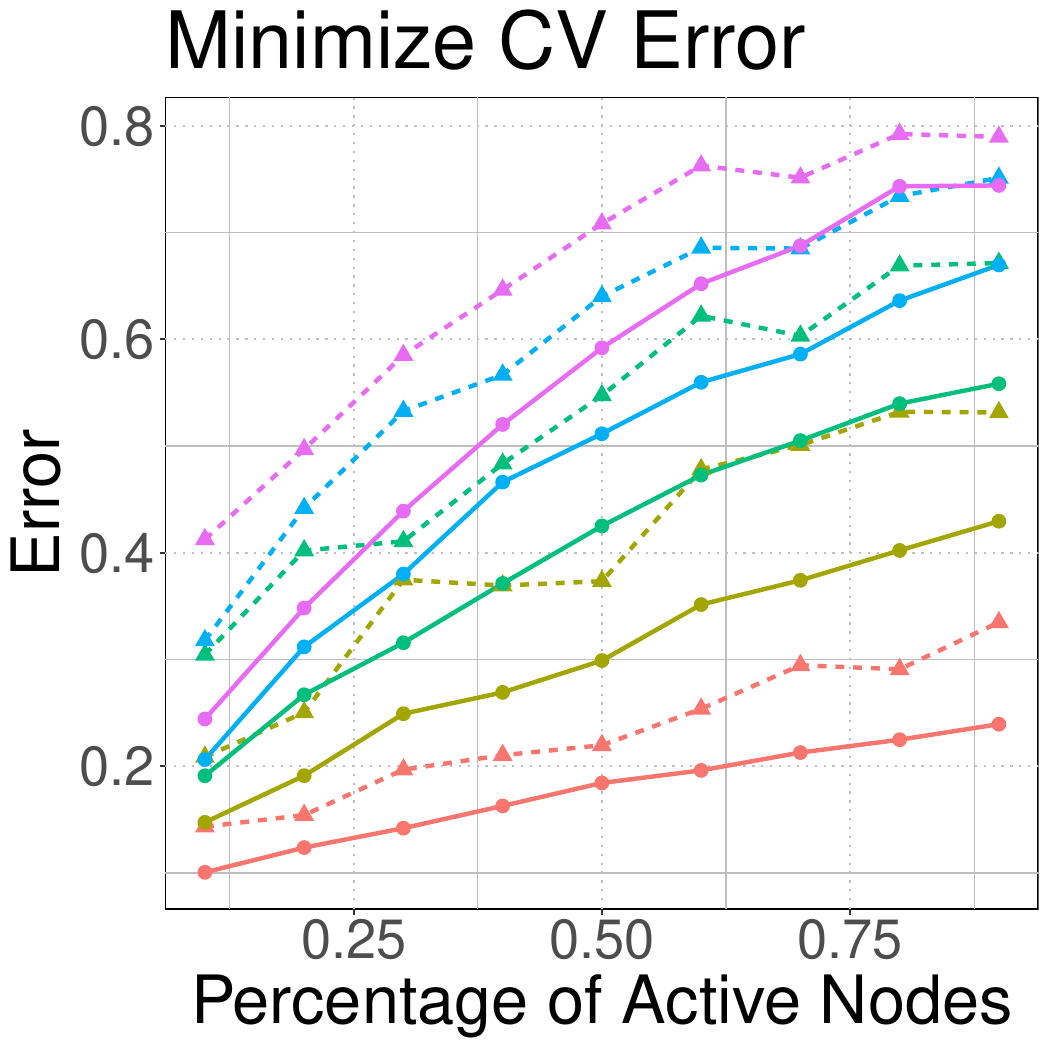}
        \caption{Linear trend filtering ($k=1$)}
     \end{subfigure}
\caption{ We vary the size of jumps at breakpoints (colors) along with the percentage of active nodes in the graph, and compare graph cross-validation against ordinary cross-validation in each case (with $L_{1}$ penalty, $\sigma^{2} = 1$, $10$ folds). The relative performance of graph cross-validation (dotted) compared to ordinary cross-validation (solid) increases 
with both the size of jumps and number of breakpoints, indicating that less smooth trends benefit the most from using graph fission to tune $\lambda$.}
    \label{fig:graph_fission_cv_results}
\end{figure*}

To illustrate this disadvantage, we vary two parameters during simulation. The first is the percentage of nodes that are allowed to be active. The second is the overall size of discontinuities that are allowed at each of the active nodes. Results are shown in \cref{fig:graph_fission_cv_results}. When the structural trend is smooth (i.e. fewer active nodes and smaller jump sizes), there is little to no difference between the approaches, but trends that are volatile benefit significantly from using graph fission.

In the Appendix, we repeat this experiment with misspecified errors (drawn from $t$, skewed normal, and Laplace distributions), and the results are nearly identical compared to the correctly specified case. These results suggest that a central limit theorem in the vein of \cite{austern2020asymptotics} may hold for graph cross-validation, but we leave theoretical guarantees an area of future investigation. 

\section{Inference After Trend Estimation} \label{sec:inference_after_trend_estimation}

After selecting $\lambda$ via graph cross-validation, an analyst may wish to perform inference on the structural trend in addition to having a single point estimate. Unfortunately, even for fixed $\lambda$, out-of-the-box inferential procedures are not available because the active set of variables are chosen adaptively based on the data. 

\cite{data_fission} provides a method for confidence interval construction in the case of univariate trend filtering, and when \cref{assumption:conv_close} holds, this general framework applies here with slight modifications. To summarize, an analyst can apply \cref{fact:data_thinning} to generate two synthetic graphs $\mathcal{G}^{\text{sel}}$ and $\mathcal{G}^{\text{inf}}$. Then, $Y^{\mathcal{G}^{\text{sel}}}$ can used to select a \emph{basis} that the analyst constrains $\hat{\beta}$ to fall into the span of. $Y^{\mathcal{G}^{\text{inf}}}$ is then used to calculate $\hat{\beta}$ by projecting the held-out graph onto the chosen basis. Since choosing a basis was done independently of the projection step, standard methods for producing confidence intervals will have proper coverage. 

Unique complications arise in the graph setting when nuisance parameters need to be estimated from the data prior to applying \cref{fact:data_thinning}. Focusing on the Gaussian setting, prior work \citep{data_fission,neufeld2023data} only provides inference procedures with rigorous guarantees when the error variance is known a priori or when the dimension of the problem is fixed. Since nodes and edges come online in tandem, the dimension of the penalty matrix will increase with $n$, invalidating the validity of these procedures. Some asymptotic guarantees in a high dimensional setting are explored in  \cite{rasines2021splitting}, but this work still relies on having access to consistent estimates of the error variance which is an assumption that tends not to hold in practice.  

We first discuss how the framework of \cite{data_fission} can be applied to this problem for the case of known nuisance parameters in \cref{sec:ste_known}. We then extend this methodology to the Gaussian case when the error variance needs to be estimated in \cref{sec:ste_unknown}.

\subsection{Inference Under \cref{assumption:conv_close}} \label{sec:ste_known}
Let $Y \sim P_{\theta}$ for some convolution closed distribution in parameter space $\Theta$ with corresponding density function $p(y,\theta)$. At the selection stage, we recommend using $L_{1}$ penalties to select a model because the basis structure of these estimates is well understood --- see \cref{alg:basis_construction} for an explicit formula for extracting a basis from a trend fit using an $L_{1}$ penalty. That said, because $\mathcal{G}^{\text{sel}}$  and $\mathcal{G}^{\text{inf}}$ are independent, the analyst is free to choose a basis in a completely arbitrary way if desired. For instance, one can use an $L_{1}$ penalty tuned with graph cross-validation as an initial step, but then manually adjust the basis via visual inspection or expert judgement. This contrasts with other methods for post-selection inference on generalized lasso problems \citep{chen2021powerful,10.1214/17-EJS1363} that are only valid when the analyst commits to using an $L_{1}$ penalty in a deterministic way. 

\begin{algorithm}[tb]
\begin{algorithmic}
   \REQUIRE{Fitted trend $\hat{\beta}$, order of penalty matrix ($k$), graph Laplacian matrix ($L$)} 
    \IF{$k$ is even}
       \STATE $C \gets L^{\frac{k}{2}} \hat{\beta}$
       \STATE Identify unique values of $C$, denoted $c_{1},...,c_{\ell}$. 
    \FOR{$t = 1,2,...,\ell$}
        \STATE Let $c_{t}$ be a vector with an entry of $1$ whenever a row of $C$ is equal to $c_{t}$ and $0$ otherwise.
    \ENDFOR
    \STATE $B \gets \left(L^{\dagger}\right)^{\frac{k}{2}} \begin{bmatrix}c_{1}^{T} & ... & c_{\ell}^{T}\\
    \end{bmatrix}$
    \ELSE
        \STATE $C \gets L^{\frac{k+1}{2}} \hat{\beta}$
       \STATE Identify $A \subseteq\{1, \ldots n\}$ corresponding to the non-zero rows of $C$. Let $B$ be $\left(L^{\dagger}\right)^{\frac{k+1}{2}}$ with only the columns corresponding to $A$ included. 
    \ENDIF
\end{algorithmic}
$B \gets \begin{bmatrix}1 & B\\
    \end{bmatrix}$
   \caption{Basis construction for $L_{1}$ penalties}
   \label{alg:basis_construction}
\end{algorithm}

\begin{lemma}
For a graph $\mathcal{G}$ and corresponding Laplacian matrix $L$, let $\hat{\beta}$ be the solution of~\eqref{eqn:opt} with $k \in \mathbb{N}$ and $D(\beta) :=\lambda \norm{\Delta^{(k)} \beta}_{1}$. Let $B$ be the output of \cref{alg:basis_construction} using $\hat{\beta}$, $k$, and $L$ as inputs. Then $\hat{\beta}$ is contained in the column span of $B$.
\end{lemma}

After selecting a basis $B \in \mathbb{R}^{m}$ using $Y^{\mathcal{G}^{\text{sel}}}$, solve for 
$$\hat{\gamma} := \argmin_{\gamma \in \mathbb{R}^{m}}\left(\sum_{i=1}^{n} p(y_{i}^{\mathcal{G}^{\text{inf}}},(1-\tau)\gamma^{T}b_{i})\right),$$
where $b_{i}$ denotes the $i$-th row of $B$. Our estimate for $\theta$ in this context is simply $B\hat{\gamma}$.  The target for inference then becomes $B\gamma$ where $\gamma := \argmin_{\gamma} D_{\mathrm{KL}}(P_{\theta} \| P_{B \gamma}) $, is the projection parameter which minimizes the KL distance between the true distribution and the working model. Inference can then be performed on this parameter using sandwich estimators for variance, as described in Theorem~4 of \cite{data_fission}.

\begin{figure}  
\centering
    \begin{subfigure}[t]{0.4\linewidth}
         \centering
         \includegraphics[width=\textwidth]{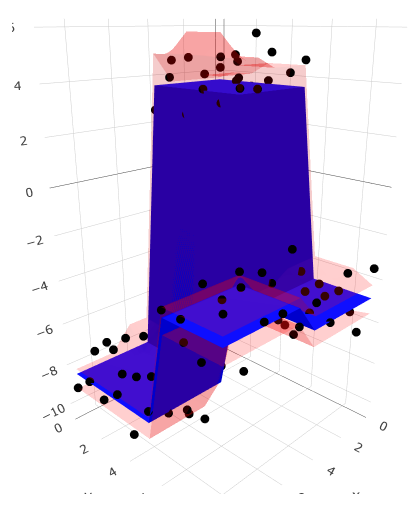}
        \caption{$k=0$.}
     \end{subfigure}
    \begin{subfigure}[t]{0.4\linewidth}
         \centering
         \includegraphics[width=\textwidth]{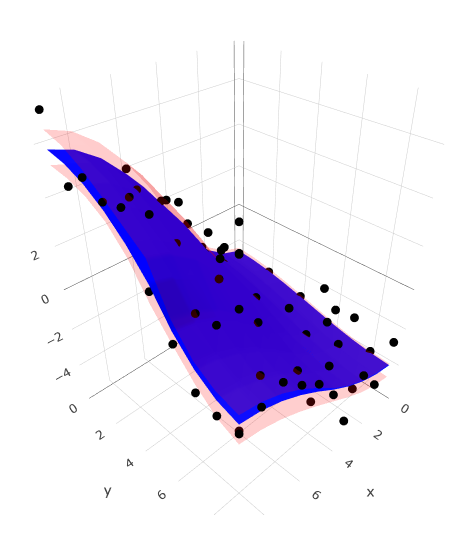}
        \caption{$k=2$.}
     \end{subfigure}
\caption{Examples of confidence intervals (red) constructed from \cref{thm:robust-ci} for two example runs. Ground truth (blue) generated from piecewise constant (left) or quadratic (right) bases. }
    \label{fig:example_graph_CI}
\end{figure}

\subsection{Inference with Nuisance Parameters} \label{sec:ste_unknown} 

One key assumption that we seek to relax is that the distribution of errors is known \emph{exactly}. In particular, \cref{fact:data_thinning} requires knowledge of $\sigma^{2}$ for Gaussian data which is unlikely in application. Instead of using \cref{fact:data_thinning}, we instead work within the \textbf{P2} regime defined in \cref{rmk:P2} which weakens the requirement that $Y^{\mathcal{G}^{\text{inf}}} \independent Y^{\mathcal{G}^{\text{sel}}}$ and only requires that $Y^{\mathcal{G}^{\text{inf}}} | Y^{\mathcal{G}^{\text{sel}}}$ have a tractable distribution. This allows us to consider errors of the form $\epsilon_{i} \sim N(0, \sigma^{2})$ where $\sigma$ is unknown.

To this end, we construct $Y^{\mathcal{G}^{\text{sel}}} = Y + Z$ by adding user-generated noise $Z \sim N\left(0,\sigma_{0}^{2} I_{n}\right)$ for an arbitrary choice of $\sigma_{0}$ and let $Y^{\mathcal{G}^{\text{inf}}} := Y$. After selecting a basis using $Y^{\mathcal{G}^{\text{sel}}}$, we base inference on the conditional distribution of $Y | Y^{\mathcal{G}^{\text{sel}}}$. Let $\tau:= \frac{\mathcal{I}_{Y^{\mathcal{G}^{\text{sel}}}}(\mu) }{\mathcal{I}_{Y}(\mu) } = \frac{\sigma^{2}}{\sigma^{2} + \sigma_{0}^{2}}$ be the proportion of the total Fisher information that is allocated to the selection step. Then, 
\begin{equation} \label{eqn:dist_sel}
    Y| Y^{\mathcal{G}^{\text{sel}}} \sim N \left(\mu (1-\tau) + \tau Y^{\mathcal{G}^{\text{sel}}} , \sigma^{2}(1- \tau) I_{n}\right).
\end{equation}

Given a selected basis $B \in \R^{n \times m}$, we define the target for inference as the projection of the structural trend onto the chosen basis $B(B^{T}B)^{-1}B^{T}\mu$. In what follows, we aim to construct confidence intervals that cover any component $j$ of this projection which we label $\eta^{T}\mu := e_{j}^{T}B(B^{T}B)^{-1}B^{T}\mu$ for brevity. Since the selection of a direction to project onto is based only on $Y^{\mathcal{G}^{\text{sel}}}$, we assume $\eta = h(Y^{\mathcal{G}^{\text{sel}}})$, where $h$ is some unknown deterministic function of the data. 

\begin{figure*}  
    \centering
    \includegraphics[width=0.4\textwidth] 
    {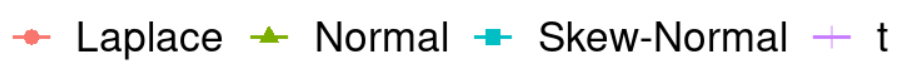}
    \newline
 \begin{subfigure}[t]{0.66\linewidth}
         \centering
         
        \includegraphics[width=0.3\textwidth]{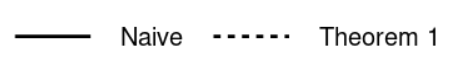}
        \newline
        \includegraphics[width=0.49\textwidth]{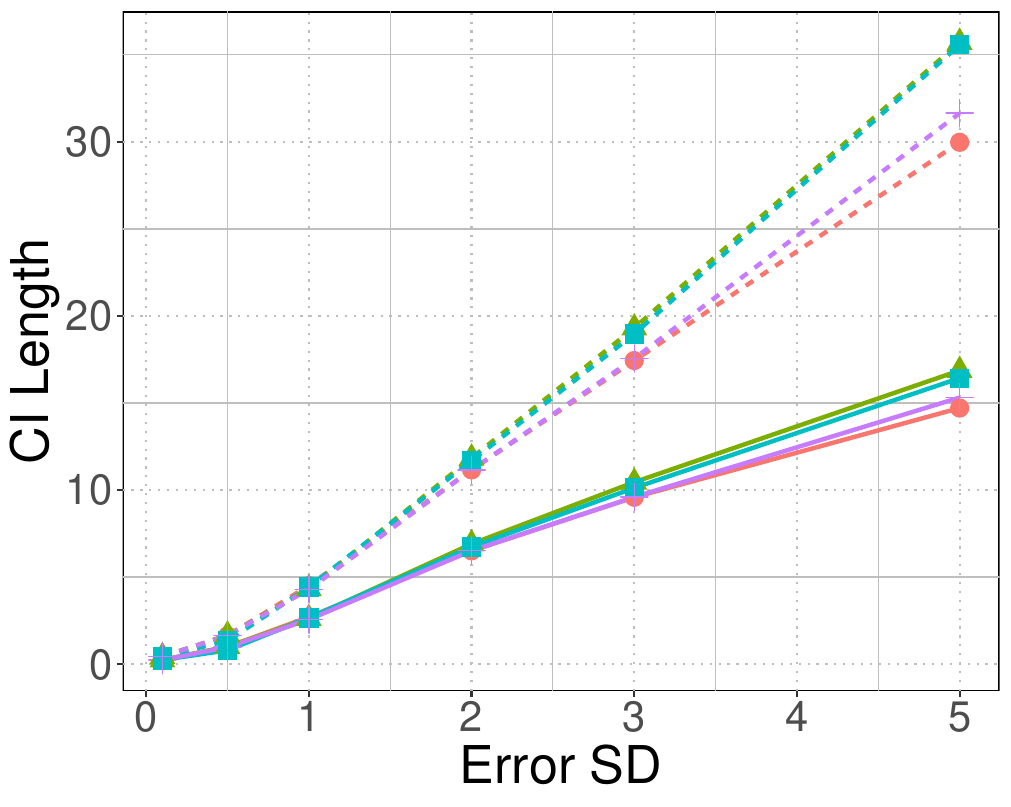}
        \includegraphics[width=0.49\textwidth]{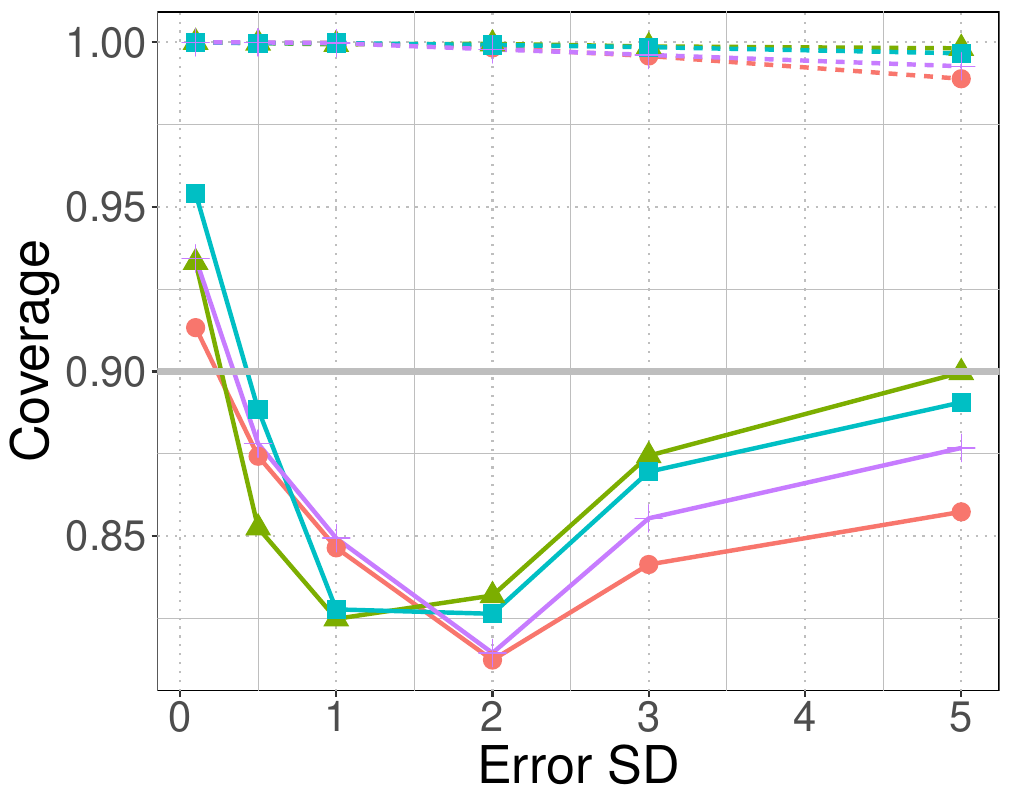}
        \caption{Confidence intervals designed using \cref{thm:robust-ci} (dashed), compared with naive intervals (solid) constructed under the assumption that $\hat{\sigma}$ is consistent.}
     \end{subfigure}
    \begin{subfigure}[t]{0.32\linewidth}
                 \centering
               \includegraphics[width=0.3\textwidth]{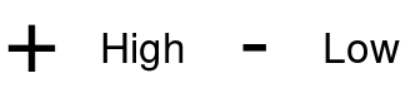}
            \includegraphics[width=\textwidth]{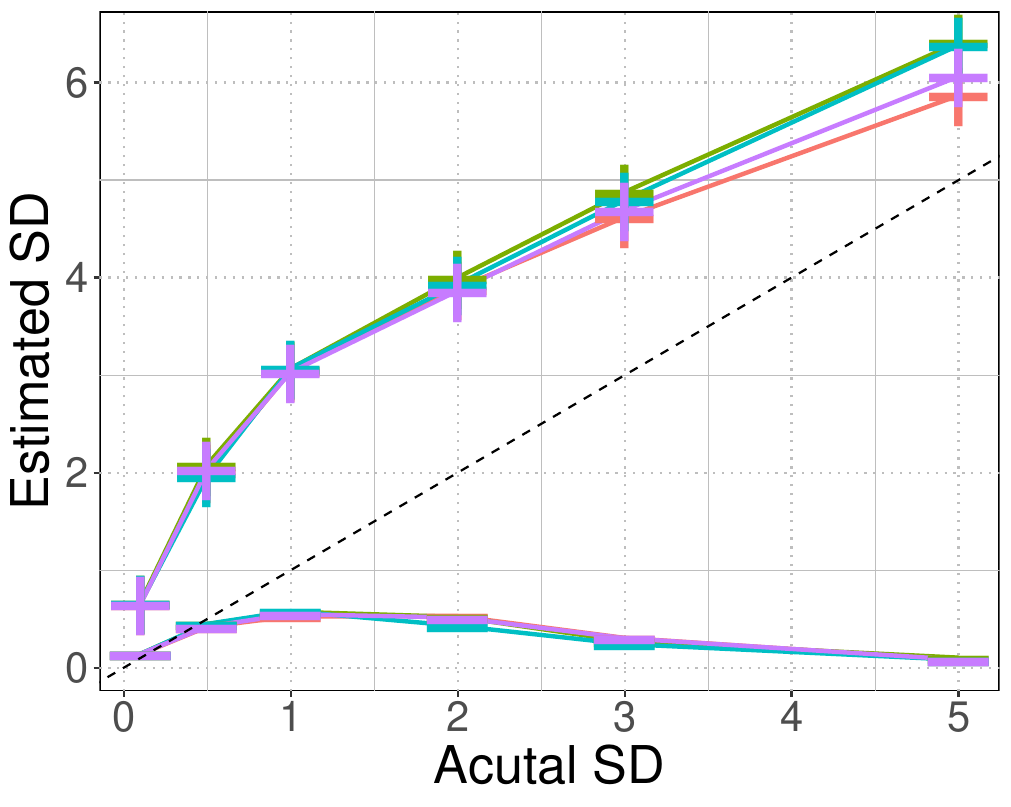}
 
            \caption{$\hat{\sigma}_{\text{low}}$ (dashed) and $\hat{\sigma}_{\text{high}}$ (solid) as a function of the the true $\sigma$. }
     \end{subfigure}

    \caption{Confidence interval length and coverage constructed for linear trend filtering ($k=1$). Intervals designed using \cref{thm:robust-ci} ensure adequate coverage, while naively constructed intervals undercover for target $1- \alpha$ = 0.9. The conservatism of the confidence intervals from \cref{thm:robust-ci} are driven by the difference between conservative and anti-conservative estimates of $\sigma$ (rightmost graph).}
     \label{fig:experiments_CI_robust}

\end{figure*}

Under these assumptions, we can create a pivot to construct confidence intervals using \cref{prop:pivot} below. 
\begin{proposition} \label{prop:pivot}
Assume $Y \sim N\left(\mu, \sigma^{2}I_{n}\right)$ and $Y^{\mathcal{G}^{\text{sel}}} = Y + Z$ where $Z \sim N(0, \sigma_{0}^{2})$ is independent of $Y$. If $\eta =h(Y^{\mathcal{G}^{\text{sel}}})$ for some deterministic function $h$, then
\begin{equation*} \label{eqn:pivot}
\frac{\sqrt{1-\tau}}{\sigma \norm{\eta}_{2}}\left( \frac{\eta^{T} Y - \tau\eta^{T} Y^{\mathcal{G}^{\text{sel}}}}{1-\tau} - \eta^{T} \mu \right)| Y^{\mathcal{G}^{\text{sel}}} \sim N(0,1).
\end{equation*}
Furthermore, if $\hat{\sigma}$ is a consistent estimator of $\sigma$ conditional on $Y^{\mathcal{G}^{\text{sel}}}$ and $\hat{\tau}:= \frac{\hat{\sigma}^{2}}{\hat{\sigma}^{2} + \sigma_{0}^{2}}$ , then 
\begin{equation*} \label{eqn:pivot_est}
\frac{\sqrt{1-\hat{\tau}}}{\hat{\sigma} \norm{\eta}_{2}}\left( \frac{\eta^{T} Y - \hat{\tau}\eta^{T} Y^{\mathcal{G}^{\text{sel}}}}{1-\hat{\tau}} - \eta^{T} \mu \right)| Y^{\mathcal{G}^{\text{sel}}} \overset{d}{\to} N(0,1).
\end{equation*}
\end{proposition}
The challenge in using this pivot is to find a consistent estimator for $\sigma$ which, as discussed, is a notoriously difficult problem in high dimensional regimes. Empirical studies \citep{hd_var_estimates} confirm that many common estimators of variance are downward biased.


A more tractable goal is to bound the true variance $\sigma^{2}$ between a conservative estimate and anti-conservative estimate and then use these bounds to construct a confidence interval, which we detail in \cref{thm:robust-ci}.
\begin{theorem} \label{thm:robust-ci}
Assume we have access to $\hat{\sigma}_{\text{high}}$, $\hat{\sigma}_{\text{low}}$ such that  $\lim_{n \rightarrow \infty} \mathbb{P} \left( \sigma^{2} \in [\hat{\sigma}_{\text{low}}^{2},\hat{\sigma}_{\text{high}}^{2}] \mid Y^{\mathcal{G}^{\text{sel}}} \right) = 1$.
Also, define:
$$\hat{\tau}_{\text{low}} = \frac{ \hat{\sigma}_{\text{low}}^{2} }{ \hat{\sigma}_{\text{low}}^{2} + \sigma_{0}^{2} } \text{, } \hat{\tau}_{\text{high}} = \frac{ \hat{\sigma}_{\text{high}}^{2} }{ \hat{\sigma}_{\text{high}}^{2} + \sigma_{0}^{2} } \text{, }$$
$$  A_{1} = \min \{ \frac{\eta^{T}Y - \hat{\tau}_{\text{low}} \eta^{T}Y^{\mathcal{G}^{\text{sel}}} }{1 - \hat{\tau}_{\text{low}}},\frac{\eta^{T}Y - \hat{\tau}_{\text{high}} \eta^{T}Y^{\mathcal{G}^{\text{sel}}} }{1 - \hat{\tau}_{\text{high}}}  \},$$ 
$$  A_{2} = \max \{ \frac{\eta^{T}Y - \hat{\tau}_{\text{low}} \eta^{T}Y^{\mathcal{G}^{\text{sel}}} }{1 - \hat{\tau}_{\text{low}}},\frac{\eta^{T}Y - \hat{\tau}_{\text{high}} \eta^{T}Y^{\mathcal{G}^{\text{sel}}} }{1 - \hat{\tau}_{\text{high}}}  \}.$$ 
Then, an asymptotic $1-\alpha$ CI for $\eta^{T} \mu$ is given by:
$$ C_{1-\alpha} := \left[ A_{1} - z_{\alpha/2}\frac{\norm{\eta}_{2} \hat{\sigma}_{\text{high}}}{\sqrt{1-\hat{\tau}_{\text{high}}}}, A_{2} + z_{\alpha/2}\frac{\norm{\eta}_{2}\hat{\sigma}_{\text{high}}}{\sqrt{1-\hat{\tau}_{\text{high}}}} \right].$$
That is, $\lim_{n \rightarrow \infty} \mathbb{P} \left( \eta^{T} \mu \in C_{1-\alpha} \mid Y^{\mathcal{G}^{\text{sel}}}\right) \ge 1- \alpha$.
\end{theorem}

The price of not having consistent estimators of $\hat{\sigma}$ is that these CIs will have an irreducible length, even as $n \rightarrow \infty$. We quantify this gap explicitly in \cref{prop:CI_length}.

\begin{corollary} \label{prop:CI_length}
Assume the conditions of \cref{thm:robust-ci} hold and $\norm{\eta}_{2} \rightarrow 0$ as $n \rightarrow \infty$, $\hat{\sigma}_{\text{low}} \overset{p}{\to} \sigma_{\text{low}} $, $\hat{\sigma}_{\text{high}} \overset{p}{\to} \sigma_{\text{high}}$ for some $\sigma_{\text{high}} \ge \sigma$, and $\sigma_{\text{low}} \le \sigma$. Then, the length of $C_{1-\alpha}$ converges in probability to
$$\eta^{T} \mu \left( \frac{ \sigma^{2}_{\text{high}} - \sigma^{2}_{\text{low}} }{ \sigma^{2}} + \frac{ \sigma^{2}_{\text{high}} - \sigma^{2}_{\text{low}} }{  \sigma_{0}^{2}} - \frac{ \sigma^{2}_{\text{high}} - \sigma^{2}_{\text{low}} }{ \sigma^{2} + \sigma_{0}^{2}}   \right).$$
\end{corollary}
The upshot of this result is that the CI length is reducible by constructing more precise estimates of $\sigma$. In the case that the analyst manages to construct estimators that are indeed consistent, the CI length will converge to $0$, and the interval will cover at an approximately $1-\alpha$ level instead of being conservative. 

In the worst case, we can select $\hat{\sigma}^{2}_{\text{low}} := 0 $ and use the sample variance as an overestimate, letting $\hat{\sigma}^{2}_{\text{high}} := \frac{1}{n-1} \sum_{i=1}^{n} (y_{i} - \bar{y})^{2}$, which \cite{uniform_asymptotic} demonstrate is a provable overestimate under mild regularity conditions. However, the length of intervals constructed using such loose bounds may be undesirably wide. More intelligent estimates can be chosen using empirical studies such as \cite{hd_var_estimates}. For an overestimate, estimators in the form of \eqref{eqn:variance_estimate} where $\lambda$ is chosen via graph crass-validation and a ``one-standard error rule'' have a tendency to be conservative. For an underestimate, first choose a basis through a lasso penalty and graph cross-validation. An estimate of the the variance using the residual standard error from an OLS regression fit on the same data using this choice of basis will be downward biased due to not adjusting for the selection step. We leave rigorous guarantees for estimators of this type an open line of inquiry.

\subsection{Simulations} We construct synthetic datasets with a ground truth mean $\mu$ in exactly the same manner as \cref{sec:cross-validation}. We choose $\sigma_{0}$ using the estimator defined in \eqref{eqn:variance_estimate}. After creating $\mathcal{G}^{\text{sel}}$ and $\mathcal{G}^{\text{inf}}$, $\mathcal{G}^{\text{sel}}$ is further split into multiple graphs to select $\lambda$ via graph cross-validation and a one-standard error rule as described in \cref{sec:cross-validation}. CIs are constructed using \cref{thm:robust-ci} (note examples for a single trial with $k=0$ and $k=2$ in \cref{fig:example_graph_CI}).  For comparison, we also consider the naive approach that assumes $\sigma$ was estimated correctly using \eqref{eqn:variance_estimate} and constructs $\mathcal{G}^{\text{sel}}$ and $\mathcal{G}^{\text{inf}}$ using \cref{fact:data_thinning} without adjustment. 

We repeat these experiments over $500$ trials for $k=1$ and both correctly and incorrectly specified errors, and report the results in \cref{fig:experiments_CI_robust}. Results demonstrate that confidence intervals constructed from \cref{thm:robust-ci} have proper coverage and the high and low estimators indeed bound the true $\sigma$. Naive confidence intervals constructed assuming consistent estimates of $\hat{\sigma}$ severely undercover. Additional results detailing construction of confidence intervals for Poisson distributed count data are contained in the Appendix. 

\section{Application to NYC Taxi Data} \label{sec:real_data_example}

\begin{figure*}
 \centering
     \begin{subfigure}[t]{0.33\linewidth}
        \centering
        \includegraphics[width=\linewidth]{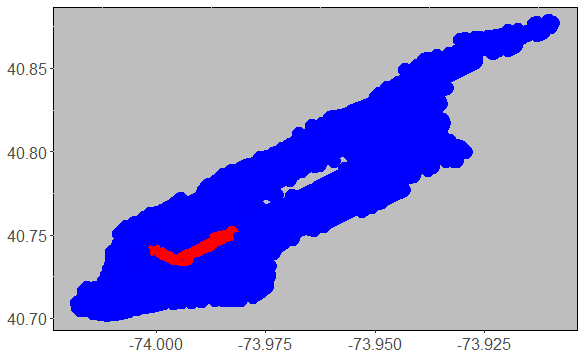}
        \caption{Ground truth}
    \end{subfigure}
    \begin{subfigure}[t]{0.33\linewidth}
        \centering
        \includegraphics[width=\linewidth]{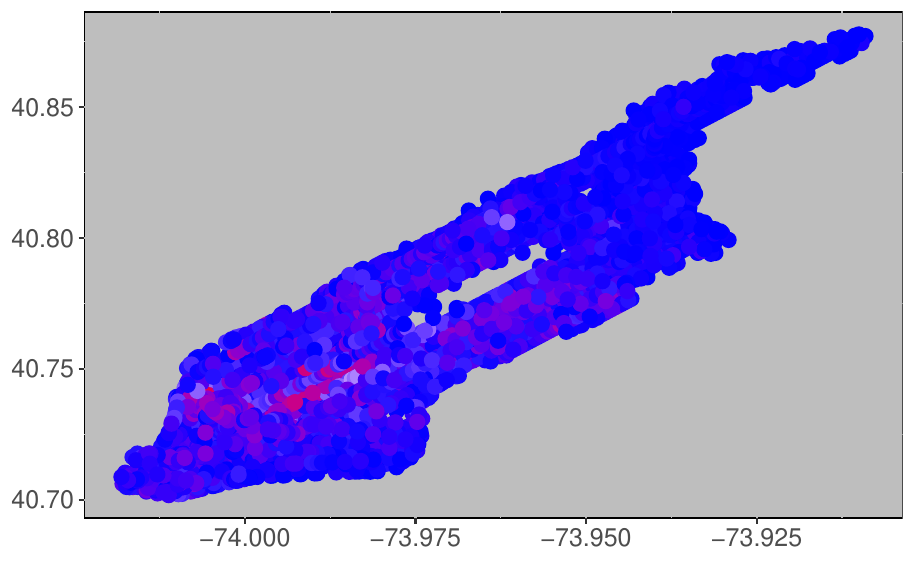}
        \caption{Raw signal}
    \end{subfigure}
    \begin{subfigure}[t]{0.33\linewidth}
        \centering
        \includegraphics[width=\linewidth]{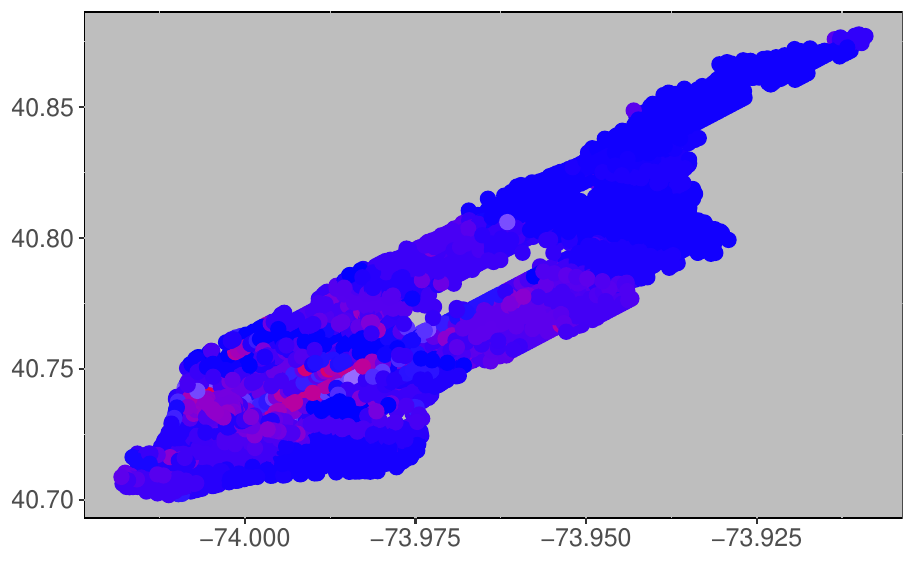}
        \caption{Lower CI}
    \end{subfigure}
    \begin{subfigure}[t]{0.33\linewidth}
        \centering
        \includegraphics[width=\linewidth]{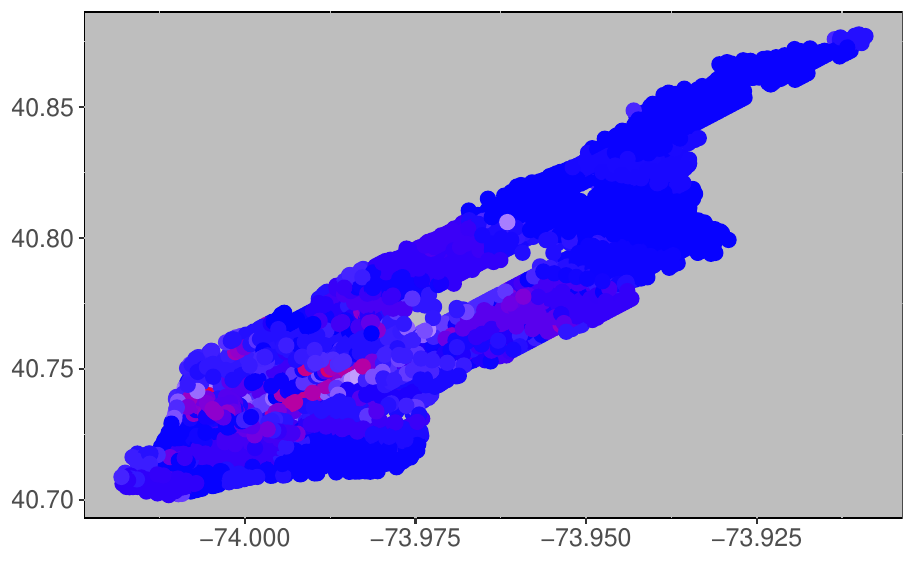}
        \caption{Upper CI}
    \end{subfigure}
    \caption{Confidence intervals around dropoffs (red means fewer and white means more than average) at each intersection. Confidence intervals correctly cover the known ground truth signal occurring due to road closures.}
   \label{fig:taxi_CI}
\end{figure*}

We conclude with a real-world application, building on an example given in \cite{graph_tf} that uses grand trend filtering to smooth a trend on the number of taxi drop-offs occurring at different intersections in Manhattan. The dataset was provided by \cite{doraiswamy}, who obtained the dataset from the NYC Taxicab and Limousine commission, and consists of nodes representing junctions (intersection between two streets) in Manhattan. Edges exist between two junctions if they are connected by a road. In total, there are 3874 nodes and 7070 edges in the dataset. 

The dataset allows us to evaluate whether confidence intervals constructed using graph trend filtering have correct coverage because during certain time periods, we have ground truth knowledge that certain roads are blocked off entirely and the number of drop-offs and pickups will be $0$ at specific junctions. One such event that we focus on is the Gay Pride parade, corresponding to a specific time period: 12:00-2:00pm on June 26, 2011. The ground truth is taken from descriptions in the news, which state that the Gay Pride parade started at 36th St. and Fifth Ave. and ended on Christopher St. in Greenwich Village.

A baseline seasonal average was constructed by averaging drop-offs  over this specific time period on the same day of each week across the nearest eight weeks. The measurement was then the difference in the amount of drop-offs during this specific time period and the seasonal average. We consider a given trend estimate as correctly reflecting the ground truth if both the lower and upper bound for the confidence intervals at these intersections are highly negative, reflecting the fact that there are significantly fewer drop-offs at these junctions due to road closures.

The results are shown in \cref{fig:taxi_CI}. The results demonstrate that graph trend filtering smooths the trend in a way that preserves the highly negative measurements along the parade route, and applying \cref{thm:robust-ci} results in confidence intervals at each junction with lower and upper limits that are also both highly negative, successfully covering the true signal. 

\section{Conclusion} \label{sec:conclude}
We extend a set of techniques for splitting information in non-$i.i.d.$ datasets  to the graph setting. We also introduce graph cross-validation, which allows for the creation of multiple folds of data with the same graph structure to select hyperparameters in graph learning problems. We apply this method to the problem of structural trend estimation on graphs, and introduce new results which enables inference when variance need to be estimated in the Gaussian case. Empirical studies show that hyperparameters chosen using graph cross-validation (nearly) minimize the risk of the estimator, and confidence intervals constructed using graph fission have correct coverage. 

We note several open questions. For applications, graph cross-validation can be used for other estimation problems on graphs, such as hyperparameter tuning for graph neural networks. 
On the theoretical side, graph fission requires an assumption of correctly specified errors. Empirical results show the procedure is nonetheless robust to modest levels of misspecification, suggesting that asymptotic guarantees are possible. 

\bibliographystyle{chicago}
\bibliography{ref}
\appendix
\section{Deferred Proofs}
\subsection{Proof of Lemma 1}
This follows from a result in \cite{graph_tf}, which we recall here. 
\begin{lemma}[Lemma 1 from \cite{graph_tf}] \label{lemma:graphtf_basis} Assume without a loss of generality that $\mathcal{G}$ is connected (otherwise the results apply to each connected component of $\mathcal{G}$). Let $D, L$ be the oriented incidence matrix and Laplacian matrix of $\mathcal{G}$. For even $k$, let $A \subseteq\{1, \ldots m\}$, and let $\mathcal{G}_{-A}$ denote the subgraph induced by removing the edges indexed by $A$ (i.e., removing edges $e_{\ell}, \ell \in A$). Let $C_1, \ldots C_s$ be the connected components of $G_{-A}$. Then
$$
\operatorname{null}\left(\Delta_{-A}^{(k+1)}\right)=\operatorname{span}\{\mathbbm{1}\}+\left(L^{\dagger}\right)^{\frac{k}{2}} \operatorname{span}\left\{\mathbbm{1}_{C_1}, \ldots , \mathbbm{1}_{C_s}\right\}
$$
where $\mathbbm{1}=(1, \ldots, 1) \in \mathbb{R}^n$, and $\mathbbm{1}_{C_1}, \ldots \mathbbm{1}_{C_s} \in \mathbb{R}^n$ are the indicator vectors over connected components. For odd $k$, let $A \subseteq\{1, \ldots n\}$. Then
$$
\operatorname{null}\left(\Delta_{-A}^{(k+1)}\right)=\operatorname{span}\{\mathbbm{1}\}+\left\{\left(L^{\dagger}\right)^{\frac{k+1}{2}} v: v_{-A}=0\right\}.
$$
\end{lemma}
Since the penalty term enforces sparsity in $\Delta^{(k)}\hat{\beta}$, we know that $\text{supp}\left(\Delta^{(k)}\hat{\beta}\right) = A$ for some active set $A \subseteq \{1,...,m\}$ which implies that $\hat{\beta} \in \text{null} \left(\Delta^{(k)}_{-A}\hat{\beta} \right)$. Working backwards, this implies that for even $k$
$$L^{\frac{k}{2}} \hat{\beta} \in L^{\frac{k}{2}} \operatorname{span}\{\mathbbm{1}\} + L^{\frac{k}{2}} \left(L^{\dagger}\right)^{\frac{k}{2}} \operatorname{span}\left\{\mathbbm{1}_{C_1}, \ldots , \mathbbm{1}_{C_s}\right\} =  \operatorname{span}\left\{\mathbbm{1}_{C_1}, \ldots , \mathbbm{1}_{C_s}\right\}.$$
So, the connected components can be identified by noting the unique values of $L^{\frac{k}{2}} \hat{\beta}$ and seeing which nodes share these values in common. For odd $k$, we have that
$$L^{\frac{k+1}{2}}  \hat{\beta} \in L^{\frac{k+1}{2}}\operatorname{span}\{\mathbbm{1}\}+L^{\frac{k+1}{2}}\left(L^{\dagger}\right)^{\frac{k+1}{2}}= v,$$
where $v$ contains $0$ at the non-zero components ($A$). However, the term $\left\{\left(L^{\dagger}\right)^{\frac{k+1}{2}} v: v_{-A}=0\right\}$ is equivalent to removing the corresponding columns from  $\left(L^{\dagger}\right)^{\frac{k+1}{2}}$ and requiring $v\in \mathbb{R}^{|A|}$ instead of $\mathbb{R}^{n}$, leading to the construction given in Algorithm 1. 
\subsection{Proof of Proposition 1}
We start by recalling that
$Y| f(Y) \sim N \left(\mu (1-\tau) + f(Y)\tau , \sigma^{2}(1- \tau) I_{n}\right)$.
By assumption $\eta$ is a deterministic function of $f(Y)$. Therefore $Y | \eta = \hat{\eta}, f(Y)$ is equal in distribution to $ Y | f(Y)$ for all $\hat{\eta}$. Thus, we will treat $\eta$ as fixed conditional on $f(Y)$. Standard properties of the multivariate normal give us that
$ \eta^{T} Y | f(Y) \sim N\left(\eta^{T}\mu + (1- \tau)\eta^{T}f(Y), \eta^{T}\eta \mu (1- \tau)\right)  $. We then rearrange terms to arrive at the pivot
$$\frac{\sqrt{1-\tau}}{\sigma \norm{\eta}_{2}}\left( \frac{\eta^{T} Y - \tau\eta^{T} f(Y)}{1-\tau} - \eta^{T} \mu \right)| f(Y) \sim N(0,1).$$
Finally, if $\hat{\sigma}$ is consistent conditional on $f(Y)$, then so is $\hat{\tau}$ by the continuous mapping theorem. We can therefore apply the continuous mapping theorem once again to conclude that 
$$
\frac{\sqrt{1-\hat{\tau}}}{\hat{\sigma} \norm{\eta}_{2}}\left( \frac{\eta^{T} Y - \hat{\tau}\eta^{T} f(Y)}{1-\hat{\tau}} - \eta^{T} \mu \right)| f(Y) \overset{d}{\to} N(0,1).
$$

\subsection{Proof of Theorem 1}
Let $E_{1} = \{ \hat{\sigma}_{\text{low}} > \sigma\}$ and $E_{2} = \{ \hat{\sigma}_{\text{high}} < \sigma\}$, and $E_{3} =\{ \hat{\sigma}_{\text{low}} < \sigma\} \bigcap \{ \sigma < \hat{\sigma}_{\text{high}} \}$. 
We then have that
\begin{align*}
\lim_{n\rightarrow \infty} \mathbb{P}\left( E_{3}| f(Y)\right) 
&= \lim_{n\rightarrow \infty} 1 - \mathbb{P} \left( E_{1} \bigcup E_{2} | f(Y)\right) \\
&=\lim_{n\rightarrow \infty} 1- \mathbb{P} \left( E_{1} | f(Y)\right) - \mathbb{P} \left( E_{2} | f(Y)\right)\\
&= 1.
\end{align*}
We already know from Proposition 1 that
$$C_{1} := \left[  \frac{\eta^{T} Y - \tau\eta^{T} f(Y)}{1-\tau} - z_{\alpha/2}\frac{\sigma \norm{\eta}_{2}}{\sqrt{1-\tau}},\frac{\eta^{T} Y - \tau \eta^{T} f(Y)}{1-\tau} +  z_{\alpha/2}\frac{\sigma\norm{\eta}_{2} }{\sqrt{1-\tau}}\right] $$
is a valid $1-\alpha$ confidence interval. Letting
$$ C_{2} := \left[ A_{1} - z_{\alpha/2}\frac{\norm{\eta}_{2} \hat{\sigma}_{\text{high}}}{\sqrt{1-\hat{\tau}_{\text{high}}}}, A_{2} + z_{\alpha/2}\frac{\norm{\eta}_{2}\hat{\sigma}_{\text{high}}}{\sqrt{1-\hat{\tau}_{\text{high}}}} \right],$$
we  note that $f(\tau) = \frac{\eta^{T} Y - \tau\eta^{T} f(Y)}{1-\tau}$ is monotonic and continuous with respect to $\tau \in [0,1)$. If $\hat{\sigma}_{\text{low}} \le \sigma \le \hat{\sigma}_{\text{high}}$, then $\hat{\tau}_{\text{high}} \le \tau \le \hat{\tau}_{\text{low}}$. We conclude by invoking the intermediate value theorem to see that $ \frac{\eta^{T} Y - \tau\eta^{T} f(Y)}{1-\tau} \in [A_{1},A_{2}]$ whenever $E_{3}$ holds.
We also note $\hat{\sigma}_{\text{high}} > \sigma$ implies that  $\frac{\hat{\sigma}_{\text{high}}}{\norm{\eta}_{2}\sqrt{1-\hat{\tau}_{\text{high}}}} \ge  \frac{\sigma }{\norm{\eta}_{2}\sqrt{1-\tau}}$ because $\tau$ monotonically increases with $\sigma$ and is bounded in $[0,1]$. 
These two arguments taken together imply that $\{\eta^{T}\mu \in \text{CI}_{1} \}\cap E_{3} \subseteq  \{\eta^{T}\mu \in \text{CI}_{2}\}\cap E_{3}$. 

Therefore, $\lim_{n\rightarrow \infty} \mathbb{P}\left( \{\eta^{T}\mu \in \text{CI}_{2} \} \right) = \lim_{n\rightarrow \infty} \mathbb{P}\left( \{\eta^{T}\mu \in \text{CI}_{2} \} \bigcap E_{3}  \right) \ge \lim_{n\rightarrow \infty} \mathbb{P}\left( \{\eta^{T}\mu \in \text{CI}_{1} \} \bigcap E_{3}  \right) = \lim_{n\rightarrow \infty} \mathbb{P}\left( \{\eta^{T}\mu \in \text{CI}_{1} \} \right) = 1- \alpha$,
concluding the proof. 
\subsection{Proof of Corollary 1}
The confidence interval length is given by $|A_{1} - A_{2}| + 2z_{\alpha/2}\frac{\norm{\eta}_{2}\hat{\sigma}_{\text{high}}}{\sqrt{1-\hat{\tau}_{\text{high}}}}$. By assumption $\norm{\eta}_{2}\rightarrow 0$ so we only have to concern ourselves with $|A_{1} - A_{2}|$. We further have that $\hat{\tau}_{\text{low}} \overset{p}{\to} \tau_{\text{low}} $, $\hat{\tau}_{\text{high}} \overset{p}{\to} \tau_{\text{high}} $, $\hat{\sigma}_{\text{low}} \overset{p}{\to} \sigma_{\text{low}} $, $\hat{\sigma}_{\text{high}} \overset{p}{\to} \sigma_{\text{high}}$ either by assumption or via the continuous mapping theorem. This implies that $A_{1} - A_{2} $ will converge in distribution to a variable that we will label $\tilde{A}$ with conditional distribution of 
\begin{multline*}
\tilde{A} | f(Y) \sim N \biggl( \eta^{T} \mu \left( \frac{ \sigma^{2}_{\text{low}} - \sigma^{2}_{\text{high}} }{ \sigma_{0}^{2}}  \right) - \eta^{T} f(Y) \left(\frac{ \sigma^{2}_{\text{low}} - \sigma^{2}_{\text{high}} }{ \sigma^{2} + \sigma_{0}^{2}} -\frac{ \sigma^{2}_{\text{low}} - \sigma^{2}_{\text{high}} }{ \sigma^{2}}\right) ,\\  \norm{\eta}_{2} \sigma^{2} (1-\tau) \left(\frac{ \sigma^{2}_{\text{low}} - \sigma^{2}_{\text{high}} }{ \sigma^{2} + \sigma_{0}^{2}} - \frac{ \sigma^{2}_{\text{low}} - \sigma^{2}_{\text{high}} }{ \sigma^{2} }  \right)I_{n} \biggr).
\end{multline*}
Because $\norm{\eta}_{2} \rightarrow 0$, the  conditional and unconditional variance of this variable will converge to $0$ and the marginal expectation will be equal to:
$$E[\tilde{A}] = E\left[E[\tilde{A}|f(Y)]\right] = \eta^{T}\mu \left( \frac{ \sigma^{2}_{\text{low}} - \sigma^{2}_{\text{high}} }{ \sigma^{2}} + \frac{ \sigma^{2}_{\text{low}} - \sigma^{2}_{\text{high}} }{  \sigma_{0}^{2}} - \frac{ \sigma^{2}_{\text{low}} - \sigma^{2}_{\text{high}} }{ \sigma^{2} + \sigma_{0}^{2}}   \right).$$
Taking everything together and then applying the continuous mapping theorem gives us the result that 
$$|A_{1} - A_{2}| \overset{p}{\to} \eta^{T} \mu \left( \frac{ \sigma^{2}_{\text{high}} - \sigma^{2}_{\text{low}} }{ \sigma^{2}} + \frac{ \sigma^{2}_{\text{high}} - \sigma^{2}_{\text{low}} }{  \sigma_{0}^{2}} - \frac{ \sigma^{2}_{\text{high}} - \sigma^{2}_{\text{low}} }{ \sigma^{2} + \sigma_{0}^{2}}   \right).$$

\section{Additional Experimental Results}
\subsection{Confidence Intervals for Poisson Data}
We repeat the experiments in Section 4, but with Poisson distributed data. We note that because no unknown parameters do not need to be estimated in this case, the methodology is more straightforward. Fact 1 can be applied directly, and confidence intervals from standard software packages generally have correct coverage of the structural trend. In particular, we use the decompositions defined in Example 2 to construct $\mathcal{G}_{\text{sel}}$ and $\mathcal{G}_{\text{inf}}$. In particular, the procedure then becomes:
\begin{enumerate}
\item Using only $\mathcal{G}_{\text{sel}}$, fit $\hat{\beta}$ as the solution to the optimization problem,  \begin{equation*}
    \hat{\beta} := \argmin_{\beta \in \mathbb{R}^{n}} \frac{1}{n}\sum_{i=1}^{n} \left( -y_{i} \beta_{i} + \exp(\beta_i) \right) + \norm{\beta}_{1}.
\end{equation*}
\item The solutions will be sparse, so construct a basis $B$ using Lemma 1 as before. 
\item Fit a new trend $\hat{\gamma}$ using any implementations of GLMs with $B$ as the set of corresponding covariates. To generate confidence intervals, we recommend using sandwich estimators of variance as described in \cite{data_fission}. For instance, those implemented in the\textsc{clubSandwich} package in R.  
\item The confidence intervals will cover the projection parameter $\gamma := \argmin_{\gamma} D_{\mathrm{KL}}(P_{\theta} \| P~_{B \gamma})$, where $P_{\theta}$ denotes the Poisson distribution with parameter $\theta$. 
\end{enumerate}

\begin{figure}  
    \centering
    \includegraphics[width=0.4\linewidth]{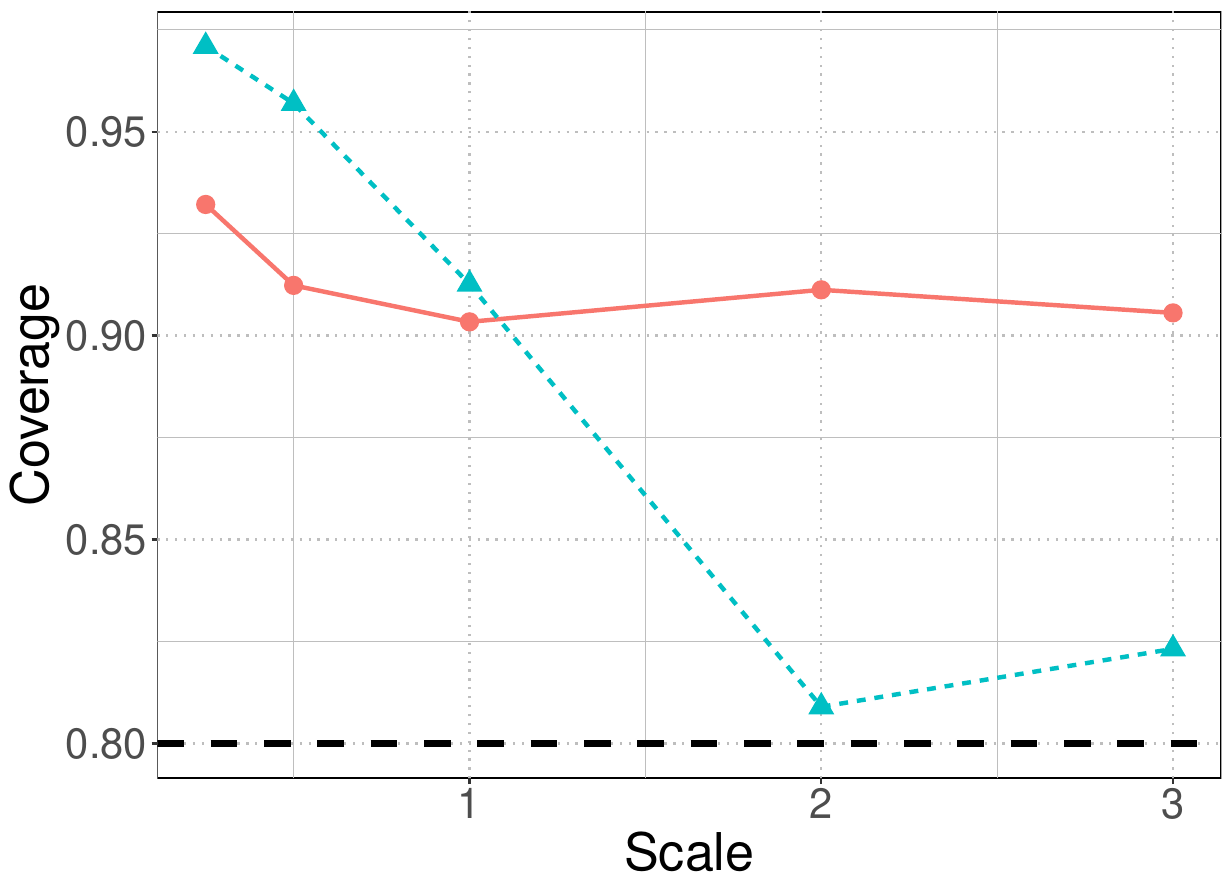}
    \includegraphics[width=0.4\linewidth]{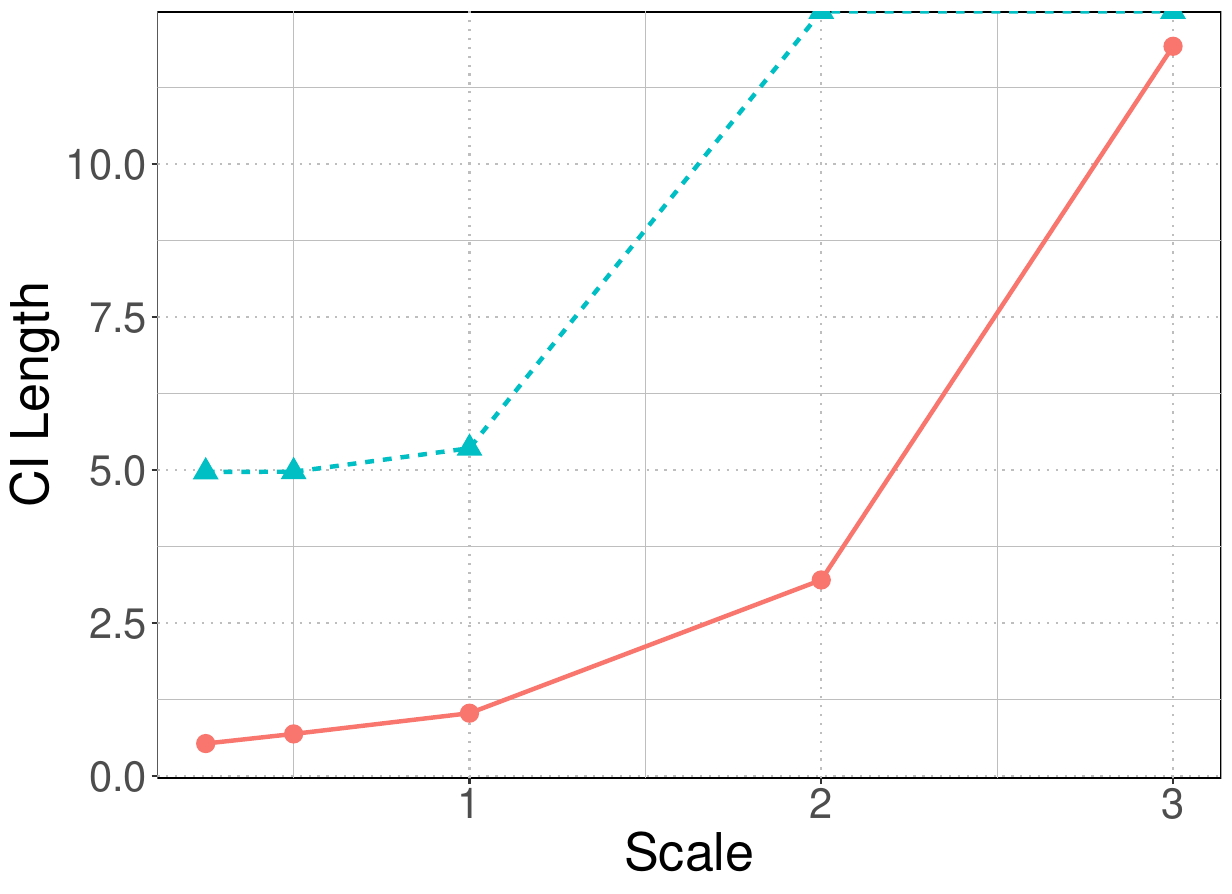}
    \caption{CI length and coverage (against target $1-\alpha$ = 0.8) as magnitude of underlying trend $\theta$ increases for Poisson-distributed data with $k=0$ (solid) and $k=1$ (dashed) graph penalties. The procedure guarantees correct coverage empirically, with confidence intervals increasing with larger $\theta$.}
     \label{fig:poisson_results}
\end{figure}

Experimental results are shown in \cref{fig:poisson_results} and are broadly consistent with the case of unknown Gaussian errors. 

\subsection{Additional Simulations for Graph Cross Validation}
\begin{figure}
\centering
\includegraphics[width=0.29\textwidth]{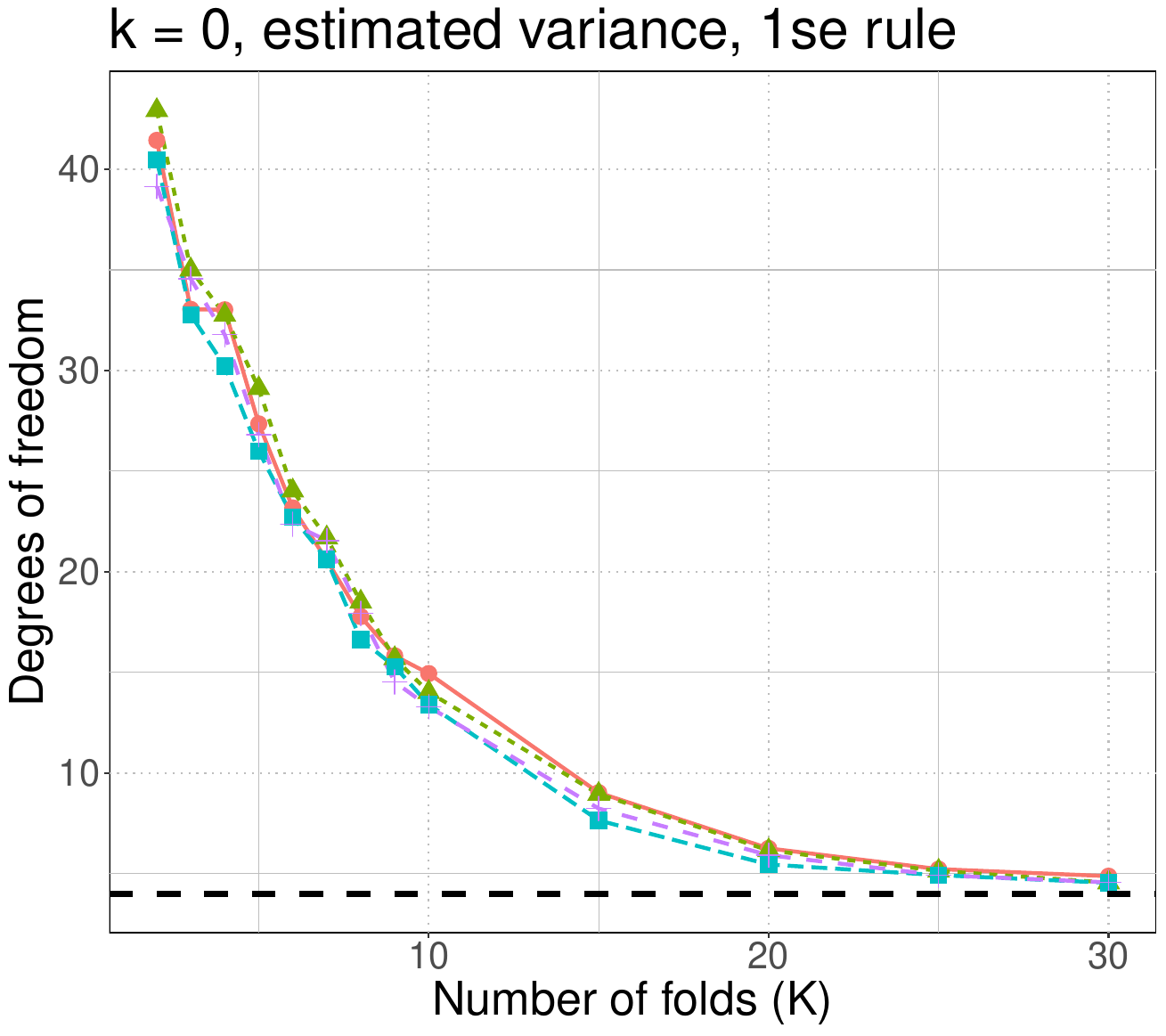}
\includegraphics[width=0.29\textwidth]{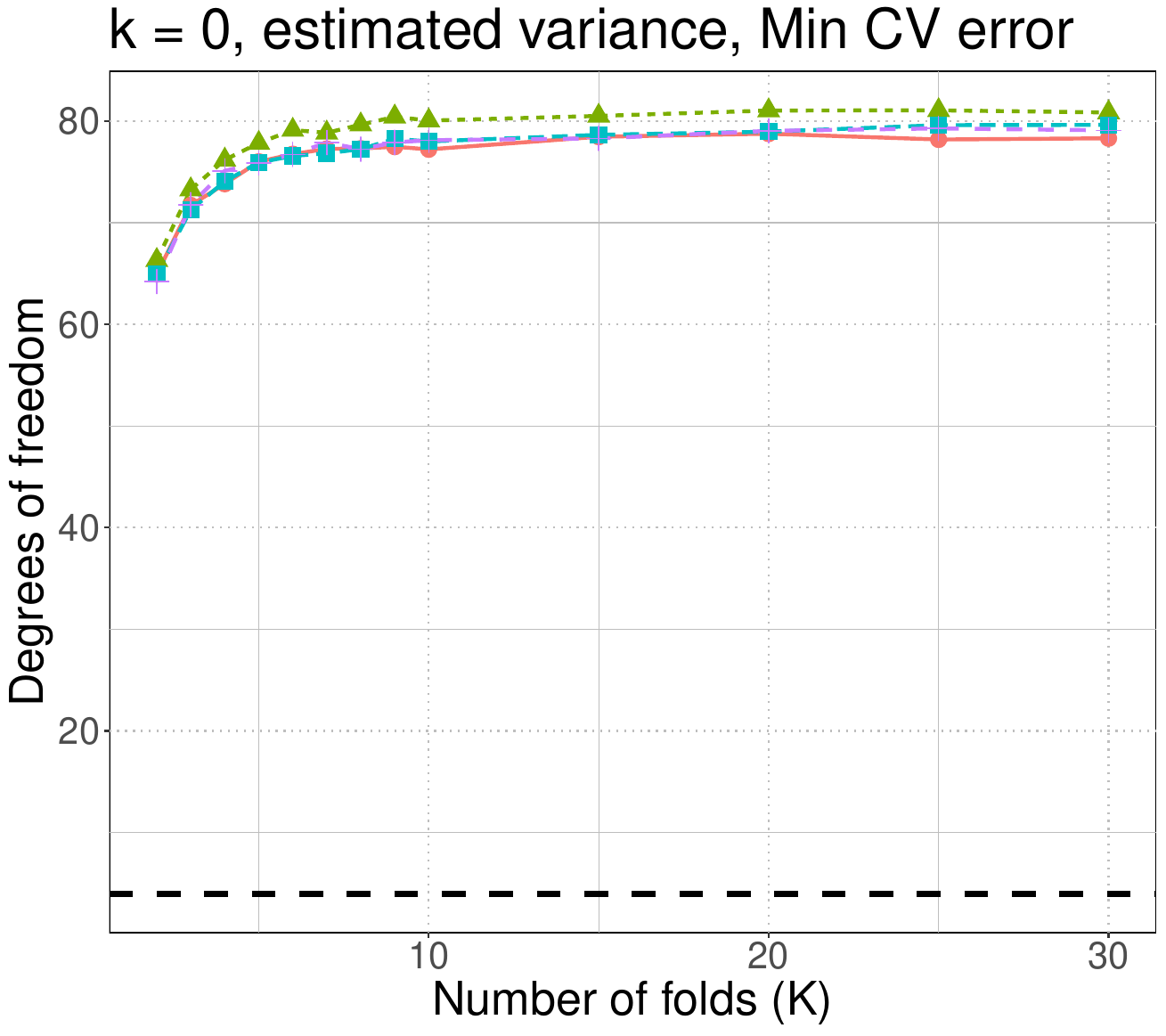}
\includegraphics[width=0.29\textwidth]{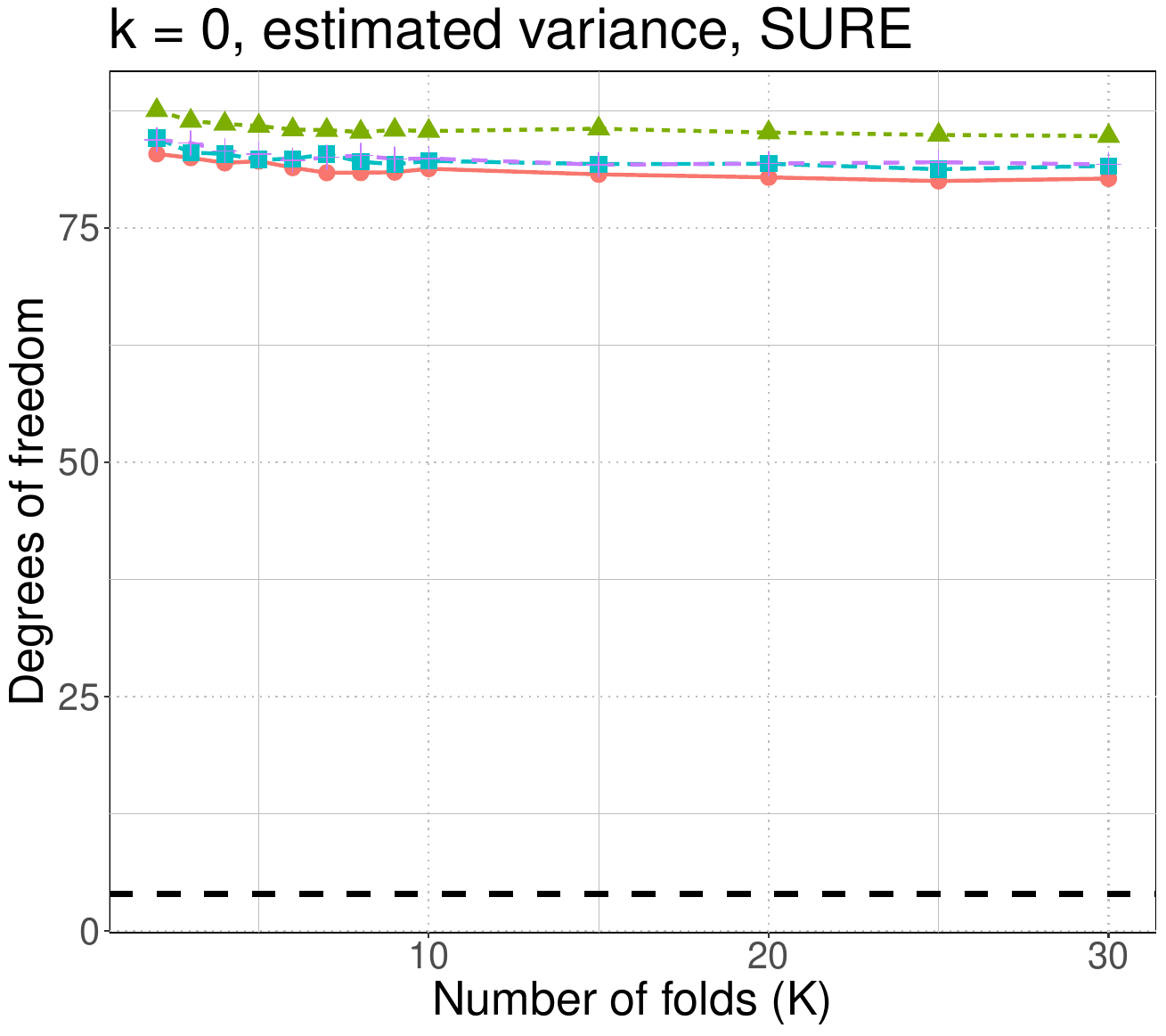}
\includegraphics[width=0.29\textwidth]{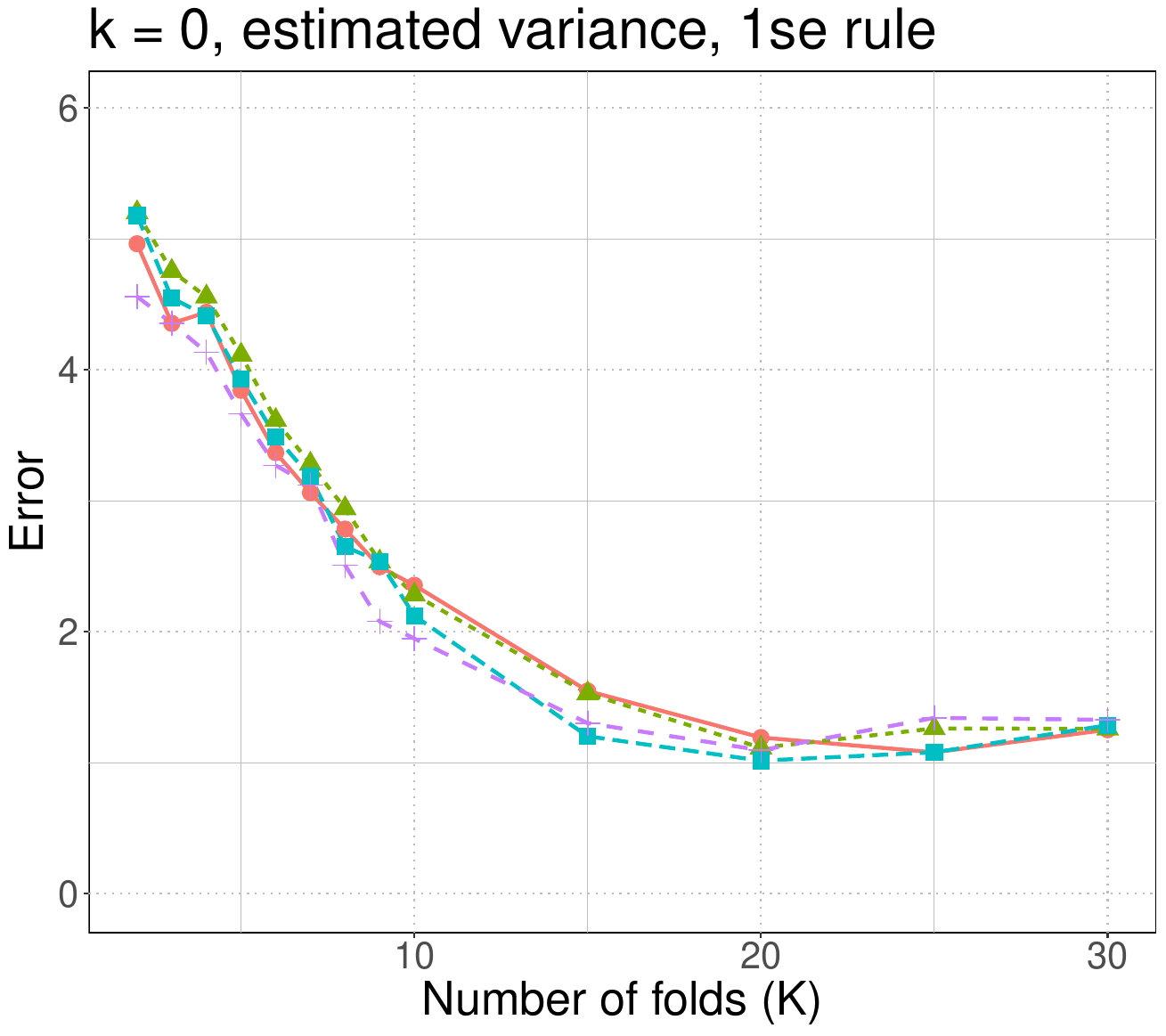}
\includegraphics[width=0.29\textwidth]{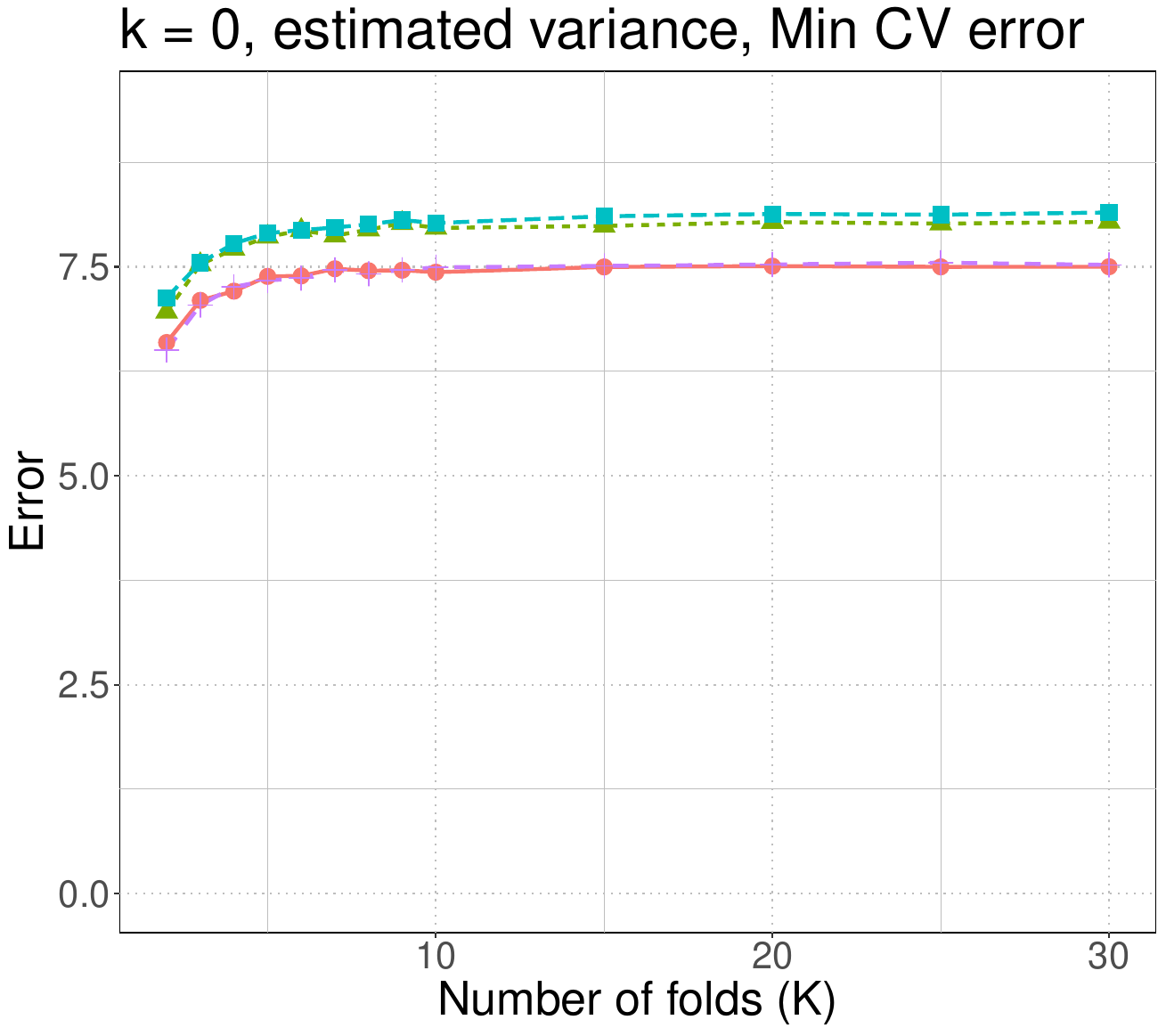}
\includegraphics[width=0.29\textwidth]{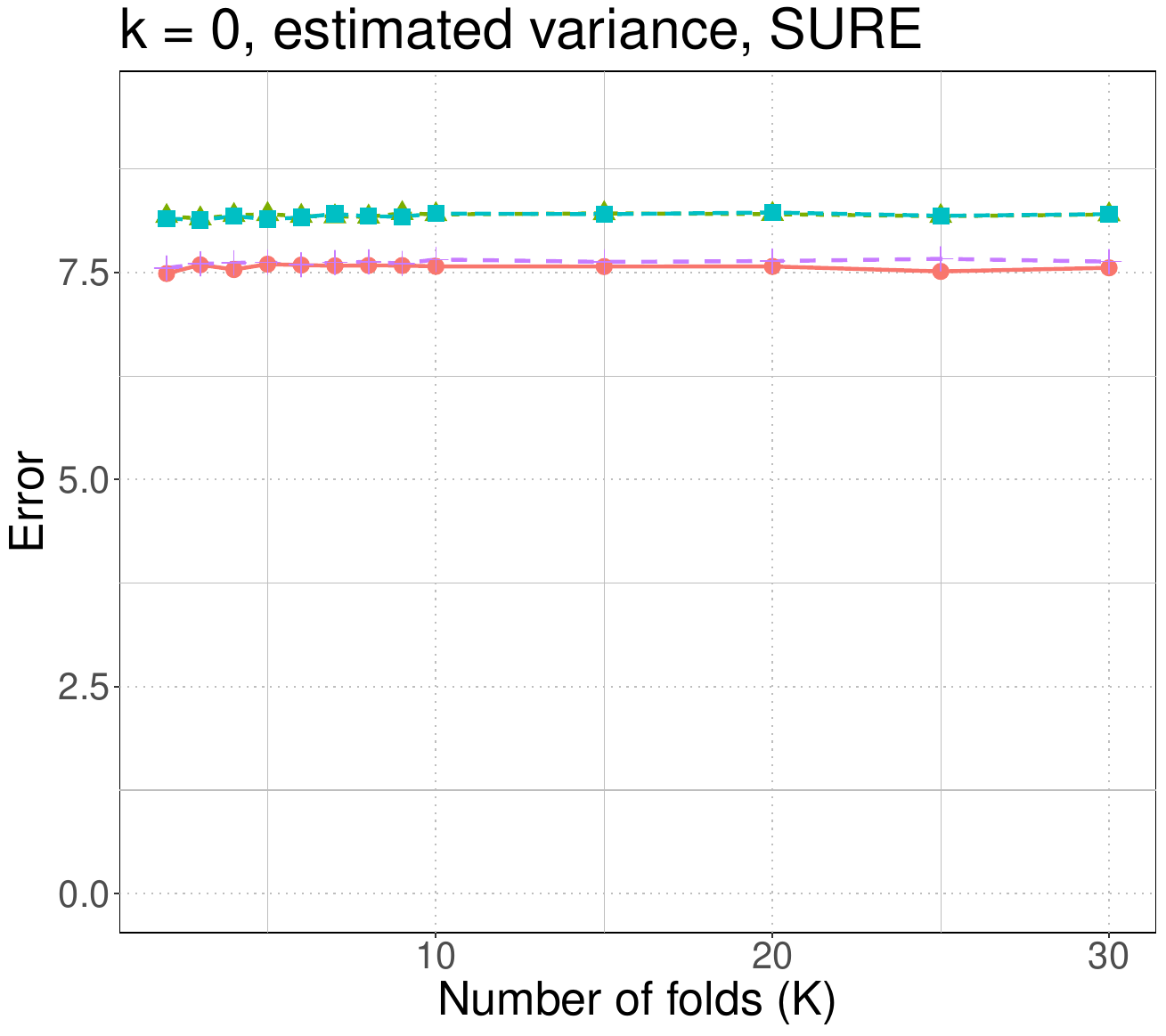}

\includegraphics[width=0.7\textwidth]{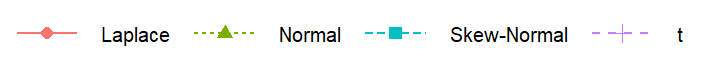}

\caption{Degrees of freedom and error when constructing $\hat{\beta}$ using piecewise constant trend filtering ($k=0$) with graph cross-validation.
Results are shown across possibly different choices for misspecification of the errors. We note that results are nearly identical across all of these permutations. This suggests that the procedure is robust to moderate levels of misspecification, and a Central Limit Theorem may apply in this case.} 
\label{fig:df_supp_k0}
\end{figure}

\begin{figure}
\centering
\includegraphics[width=0.29\textwidth]{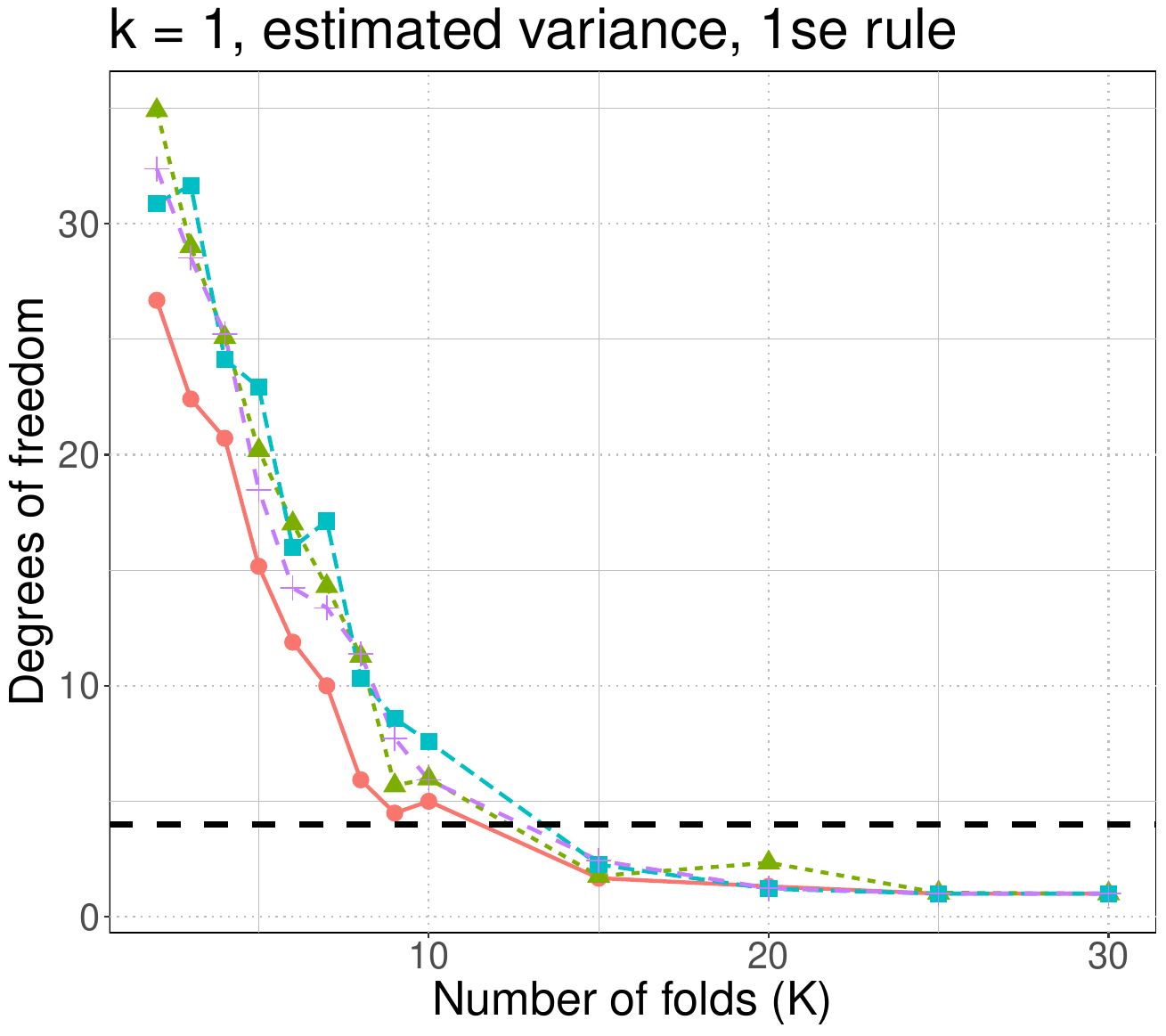}
\includegraphics[width=0.29\textwidth]{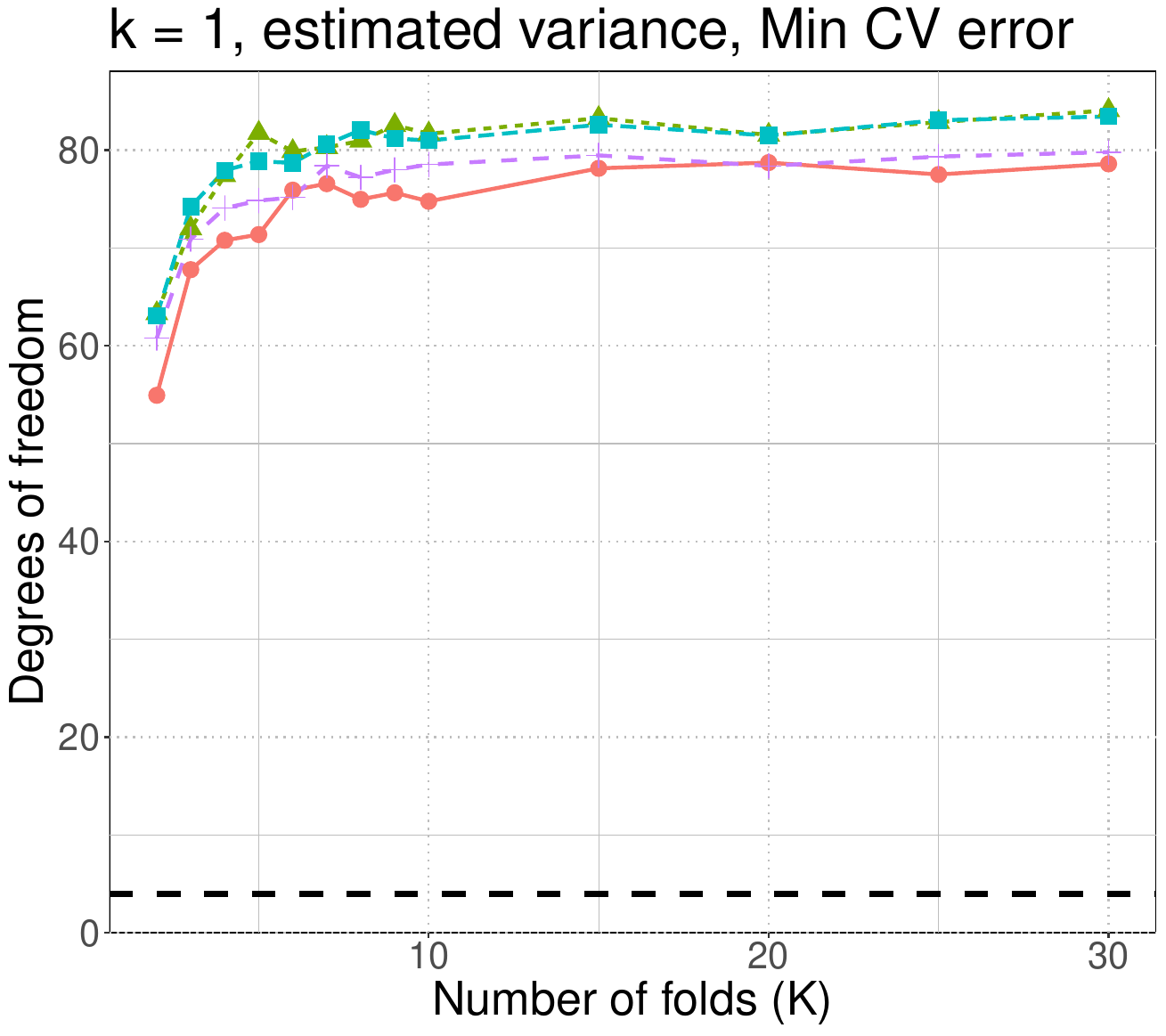}
\includegraphics[width=0.29\textwidth]{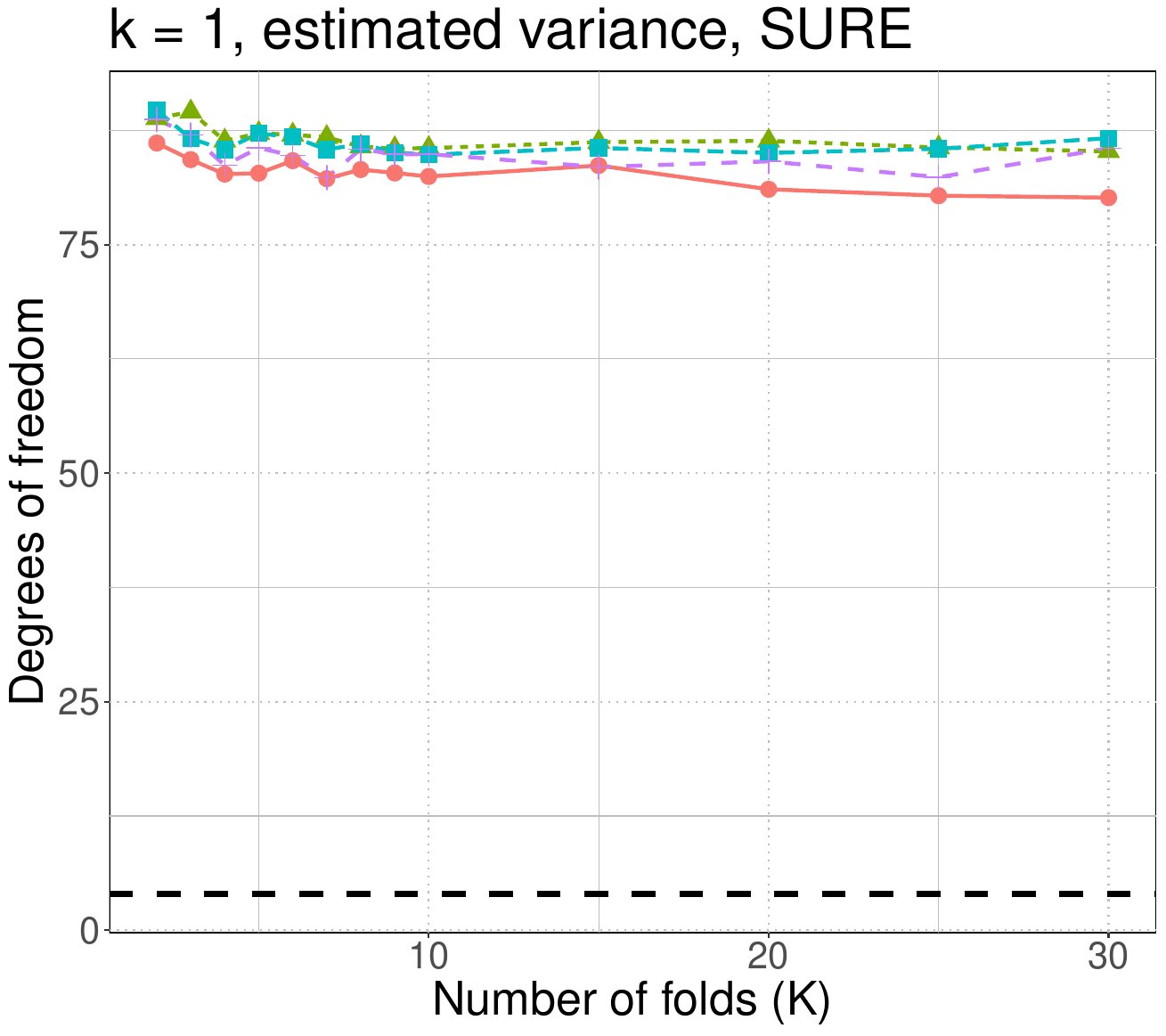}
\includegraphics[width=0.29\textwidth]{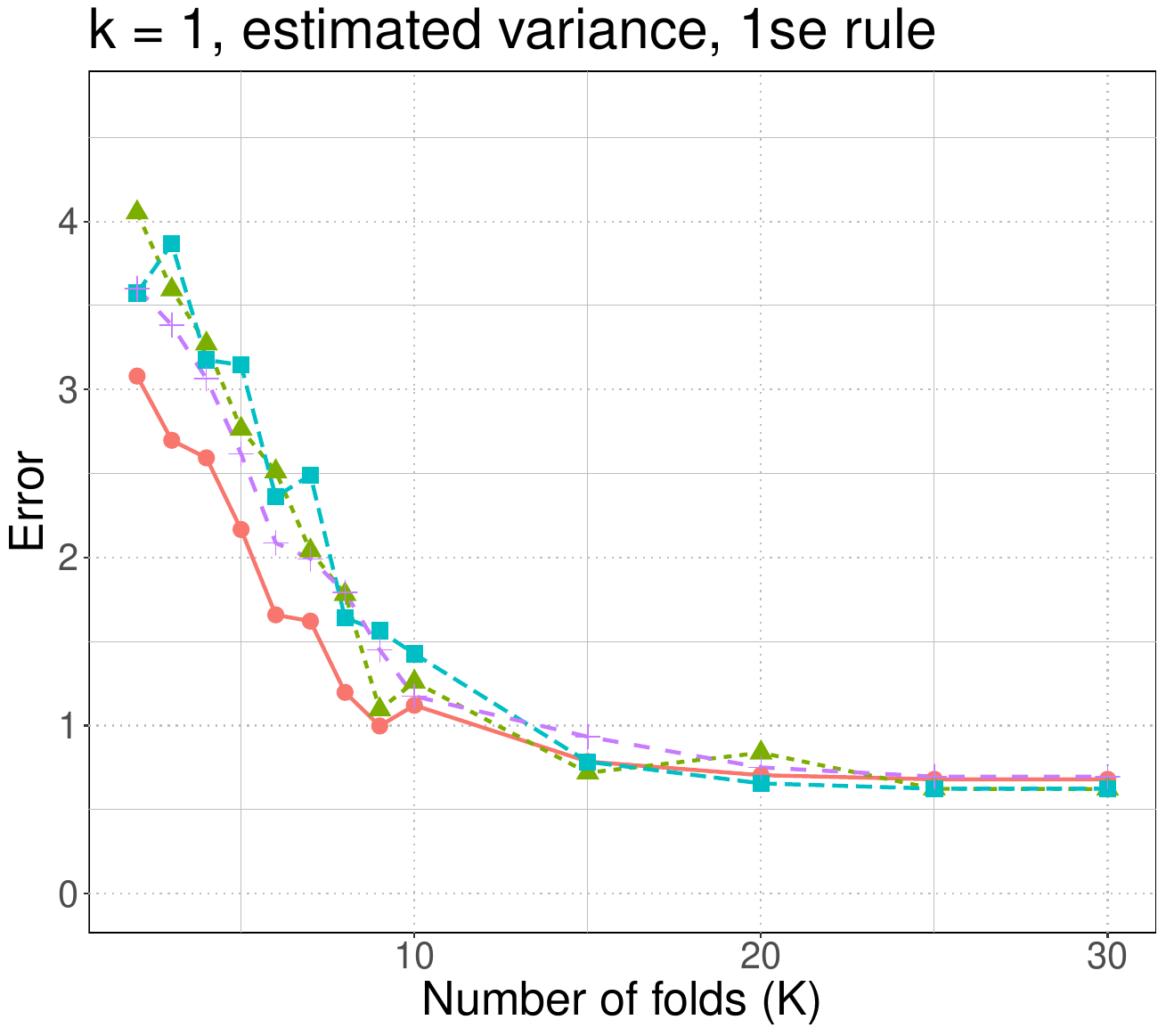}
\includegraphics[width=0.29\textwidth]{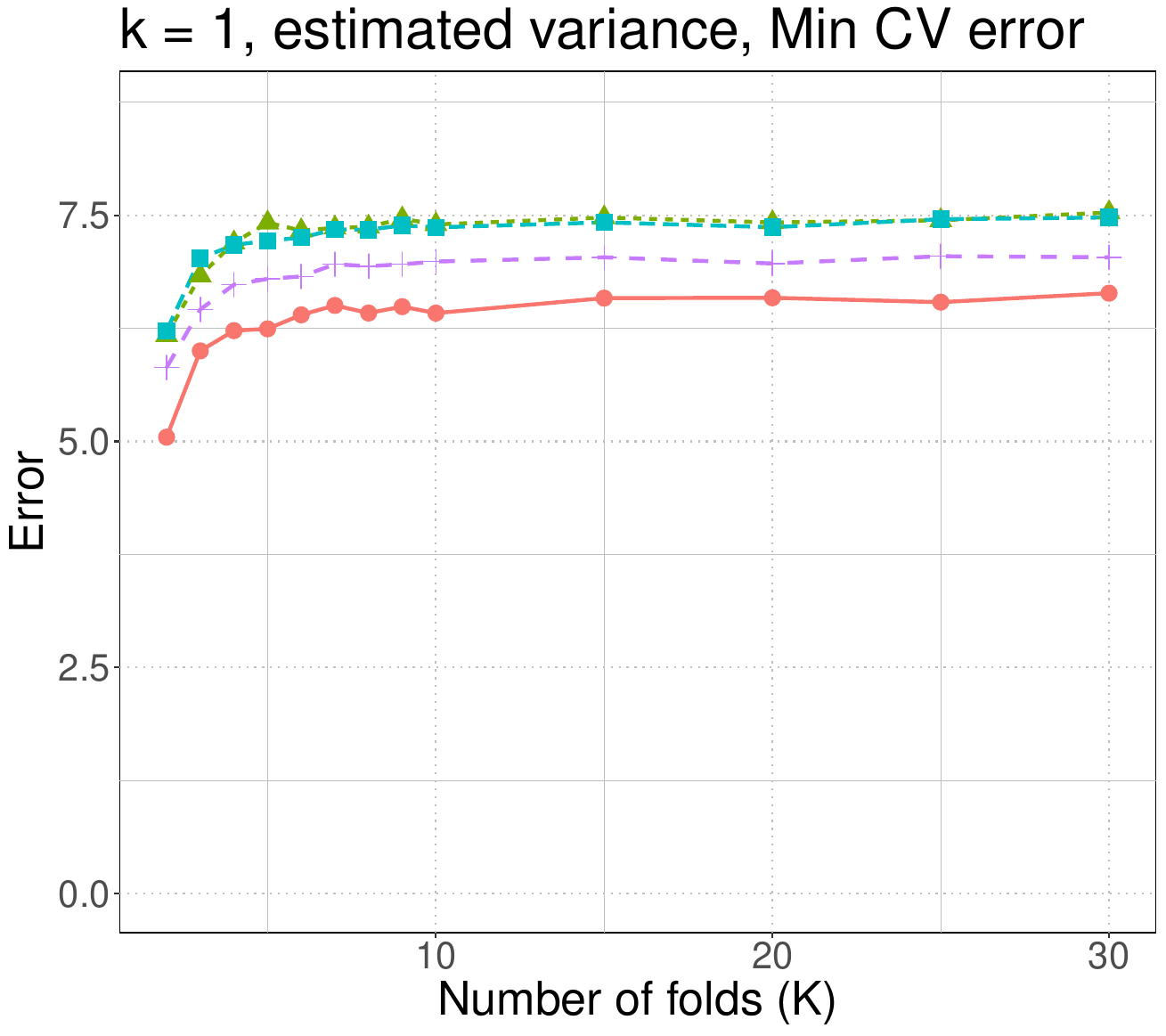}
\includegraphics[width=0.29\textwidth]{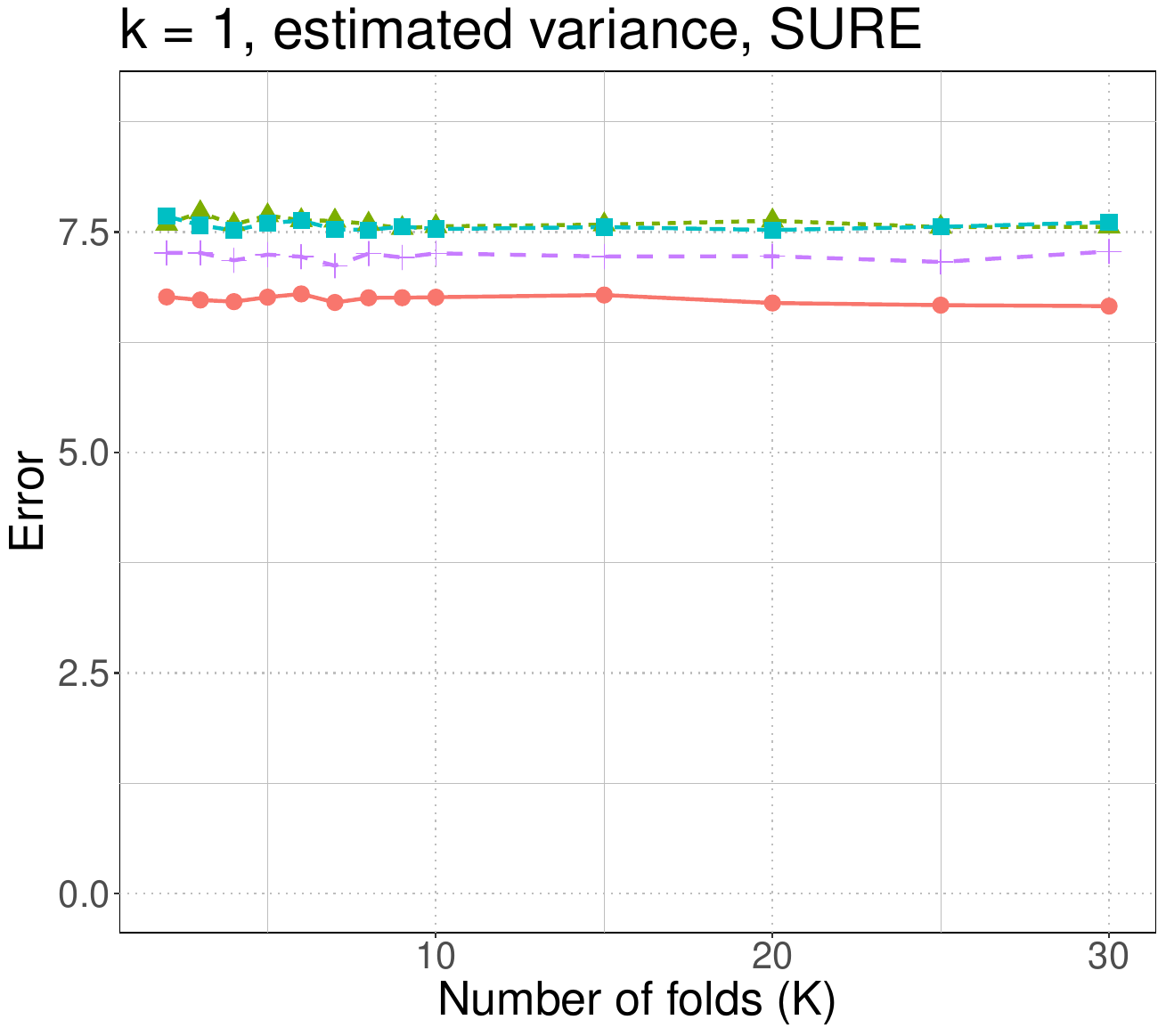}

\includegraphics[width=0.7\textwidth]{figures/legend_cv_appendix.png}

\caption{Same view as \cref{fig:df_supp_k0}, but for linear trend filtering with $k=1$. We again see similar behaviors across all error types. The spread between quantitative results across distributions is relatively larger than the case where $k=0$, though interestingly misspecified errors lead to lower risk when using cross-validation. This is an interesting area for future investigation.} 
\label{fig:df_supp_k1}
\end{figure}

We repeat the cross-validation experiments, but investigate how sensitive the methodology is to a misspecified model for the error distribution. In particular, we experiment with three different choices for the error term: Laplace, skew normal distribution with scale parameter equal to 1 and shape parameter equal to 5, and a $t$-distribution with $5$ degrees of freedom. In all cases, the error terms are also rescaled to have $0$ mean and unit variance. Results are shown for $k=0$ in \cref{fig:df_supp_k0} and $k=1$ in 
\cref{fig:df_supp_k1}. We note the trends are nearly identical across all permutations. 

\end{document}